\documentclass[%
 reprint,
 amsmath,amssymb,
 aps,
]{revtex4-1}
\usepackage{amsmath}%
\usepackage{MnSymbol}%
\usepackage{wasysym}%
\usepackage{verbatim}
\usepackage{graphicx}
\usepackage{dcolumn}
\usepackage{bm}

\newtheorem{mydef1}{Theorem}

\newtheorem{mydef2}{Lemma}

\newtheorem{mydef3}{Resource Inequality}

\newtheorem{mydef4}{Observation}
\begin{document}

\preprint{APS/123-QED}

\title{Random access codes and non-local resources }

\author{Anubhav Chaturvedi}
\email{anubhav.chaturvedi@research.iiit.ac.in}
\affiliation{Institute of  Theoretical Physics and Astrophysics, National Quantum
Information Centre, Faculty of Mathematics, Physics and Informatics, University of Gda\'nsk, Wita Stwosza 57, 80-308 Gda\'nsk, Poland}
\author{Karol Horodecki}
\affiliation{Institute of  Informatics, National Quantum
Information Centre, Faculty of Mathematics, Physics and Informatics, University of Gda\'nsk, Wita Stwosza 57, 80-308 Gda\'nsk, Poland}
\author{Marcin Pawlowski}
\affiliation{Institute of  Theoretical Physics and Astrophysics, National Quantum
Information Centre, Faculty of Mathematics, Physics and Informatics, University of Gda\'nsk, Wita Stwosza 57, 80-308 Gda\'nsk, Poland}

\date{\today}

\begin{abstract}
It is known that a PR-BOX (PR), a non-local resource and $(2\rightarrow 1)$ random access code (RAC), a functionality (wherein Alice encodes 2 bits into 1 bit message and Bob learns one of randomly chosen Alice's inputs) are equivalent under the no-signaling condition. In this work we introduce generalizations to PR and $(2\rightarrow 1)$ RAC and study their inter-convertibility.\\ 
We introduce generalizations based on the number of inputs provided to Alice, $B_n$-BOX and $(n\rightarrow 1)$ RAC. We show that a $B_n$-BOX is equivalent to a no-signaling $(n\rightarrow 1)$ RACBOX (RB). Further we introduce a signaling $(n\rightarrow 1)$ RB which cannot simulate a $B_n$-BOX. Finally to quantify the same we provide a resource inequality between $(n\rightarrow 1)$ RB and $B_n$-BOX, and show that it is saturated. As an application we prove that one requires atleast $(n-1)$ PRs supplemented with a bit of communication to win a $(n\rightarrow 1)$ RAC.\\
We further introduce generalizations based on the dimension of inputs provided to Alice and the message she sends, $B_n^d(+)$-BOX,  $B_n^d(-)$-BOX and  $(n\rightarrow 1,d)$ RAC ($d>2$). We show that no-signaling condition is not enough to enforce strict equivalence in the case of $d>2$. We introduce classes of no-signaling $(n\rightarrow 1,d)$ RB, one which can simulate $B_n^d(+)$-BOX, second which can simulate $B_n^d(-)$-BOX and third which cannot simulate either. 
Finally to quantify the same we provide a resource inequality between $(n\rightarrow 1,d)$ RB and $B_n^d(+)$-BOX, and show that it is saturated.
\end{abstract}

\pacs{Valid PACS appear here}
\maketitle

\section{Introduction}
\noindent Popescu and Rohrlich \cite{popescu1994quantum} have found that no-signaling condition allows for more non-locality (more violation of Bell inequalities) than what is allowed by quantum theory. They introduced PR-BOX (PR), the most non-local system that violate the  Clauser-Horne-Shimory-Holt (CHSH) inequality. This paved the way to the quest for information theoretic principles that restrict the violation of Bell inequalities, in particular CHSH inequality to the Tsirelson bound. Using a PR Alice and Bob could win with certainty a $(2 \rightarrow 1)$ Random Access Code (RAC). Most basic RAC is a functionality wherein Alice has two bits and is allowed to communicate only one bit to Bob, for which we shall use the notation $(2\rightarrow1)$ specifying the encoding of $2$ input bits into $1$ bit message. Bob is not allowed to communicate with Alice and guesses one of Alice's bits depending on a random choice. Within both classical and quantum theory it is not possible to win the RAC with certainty, however within the quantum theory the maximal probability of winning a RAC is higher than that within the classical theory. \\
RACs are a broad category of communication tasks which have found use in a variety of applications. For instance RAC serve as basic primitives for cryptography in classical information theory \citep{kilian1988founding,crepeau1988achieving}. Whereas in quantum scenario they were a basis of the Wiesner's first quantum protocols \cite{wiesner1983conjugate,ambainis2002dense}, semi-device independent cryptography \cite{pawlowski2011semi,chaturvedi2015security} and randomness expansion \cite{li2011semi,li2012semi}. Apart of information theoretic tasks RACs have also found application in foundations of quantum mechanics \citep{casaccino2008extrema,pawlowski2010entanglement,pawlowski2009information}. In particular the principle of information causality \cite{pawlowski2009information} was primarily based on RACs.\\
The notion of inter-convertibility between resources is fundamental to any theory of resources, for instance entanglement theory \cite{bennett1996concentrating,horodecki2002laws,brandao2008entanglement}, quantum communication theory \cite{abeyesinghe2009mother}, thermodynamics \cite{janzing2000thermodynamic,horodecki2013fundamental,brandao2013resource} or in general no-signaling theory \cite{allcock2009closed,brunner2011bound}.
The question raised in \cite{grudka2014popescu} is whether a functionality (RAC) can be used to simulate a PR. As said a PR can simulate a RACBOX (RB), an arbitrary BOX which when supplemented with one bit of communication can win a RAC with certainty. It was shown that under the no-signaling condition a RB can simulate a PR.
Thus the authors established equivalence between a no-signaling system (PR) and a functionality (no-signaling RAC). Furthermore they provided an example of a signaling RB which cannot simulate a PR. In this way a signaling resource (no signaling RB) was shown to be a weaker resource as compared to a no-signaling resource (no-signaling RB). \\
In this work we study in depth the relationship between dynamic resources RAC and static non-local resources. \\ 
The paper is divided into two parts:\\
In the first part we consider the case of bits. We introduce $B_n$-BOX a generalization of the PR with respect to number of bits supplied to Alice. We show that a $B_n$-BOX is equivalent to a no-signaling $(n\rightarrow 1)$ RB. Here the notation $(n\rightarrow 1)$ implies that Alice encodes $n$ bits into a single bit message. Further we find a bad signaling $(n\rightarrow 1)$ RB which cannot simulate the $B^n$-BOX. To quantify the same we provide a resource inequality and show that it is saturated. We show that $n-2$ $(2\rightarrow 1)$ RBs and a bit of classical communication cannot win a $(n\rightarrow 1)$ RAC. Using the equivalence we provide a protocol for winning a $(n\rightarrow 1)$ RAC using $n-1$ PRs or equivalently $n-1$ $(2\rightarrow 1)$ RBs and a single bit of classical communication.  \\
In the second part we focus on the more general case of $d$its. The case of higher dimensions is more intricate. We introduce two distinct non-local boxes $B^d_n(+)$ BOX and $B^d_n(-)$ BOX and show that these boxes can win a $(n\rightarrow 1,d)$ RAC. Here the notation $(n\rightarrow 1,d)$ specifies the encoding of $n$ $d$its into a single $d$it message where $d>2$.  However here no-signaling condition is not enough to enforce equivalence. We show using explicit examples that there exists no-signaling $(n\rightarrow 1,d)$ RB which  cannot simulate $B^d_n(+)$-BOX or $B^d_n(-)$-BOX. Yet again to quantify the same we provide a resource inequality and show that it is saturated. 

Abstract as these results may sound, they have connection to cryptography. It has been found by Wim van Dam \cite{WvD05} 
that using $2^n-1$ PR boxes and single bit of communication one can achieve $(2^n \rightarrow 1)$ RAC. It has been further noted \cite{Colbeck} that
in case of device independent key such a protocol can be used as a hacking attack. Let Alice and Bob (the honest parties) share together $n-1$ PR boxes with Eve ( an eavesdropper), where $n$ is the key length. Then upon certain wiring on their side, and leakage of single bit to Eve (e.g. by a Trojan-horse program), via the van Dam protocol Eve
is in a favorable position that she can choose to learn the particular bit of key shared by Alice and Bob. Knowing this attack, we ask if there exist a smaller box
than $n-1$ PR boxes, which does the same task, which would make such a protocol significantly more difficult to detect. We give negative answer this question:
{\it any} box which achieves the attack, is equivalent to $n-1$ PR boxes, which amounts to certain non-negligible "`memory"' inside of the Alice's and Bob's devices,
making the attack harder to perform. 

The attack invoked above can be however performed only in a world where extremal non-signaling boxes like PR box can be prepared. There are yet important reasons
for such a world do not exist. One of them is the so called Information Causality principle \cite{pawlowski-2009-461}. Indeed, it disallow not only such boxes like PR to exist, but also disallows for the access in a manner of the RAC:
\begin{equation}
\sum_i I(a_i: E |e = i) \leq 1
\end{equation}
where  $a_i$ are the bits that Alice and Bob has (the key bits), and $I$ denotes conditional mutual information (conditioned upon choice of the bit $e=i$ by Eve).
The RHS is the number of leaked bits, so that essentially the effect of RAC is suppressed: only the amount of leakage is known to Eve. One can therefore
consider the secrecy extraction not only under quantum or non-signaling eavesdropping, but also under other principles such as the Information Causality, which
we leave however for future work.

\section{\label{sec:level1}The case of bits}
In this section we provide generalization based on the number of input bits provided to Alice to $(2\rightarrow 1)$ RAC and PR. We start of with defining the resources under consideration and subsequently study their relationships.
\subsection*{$(n\rightarrow1)$ RAC, $(n\rightarrow1)$ RB and $B_n$-BOX}
\begin{figure}
\includegraphics[scale=0.5]{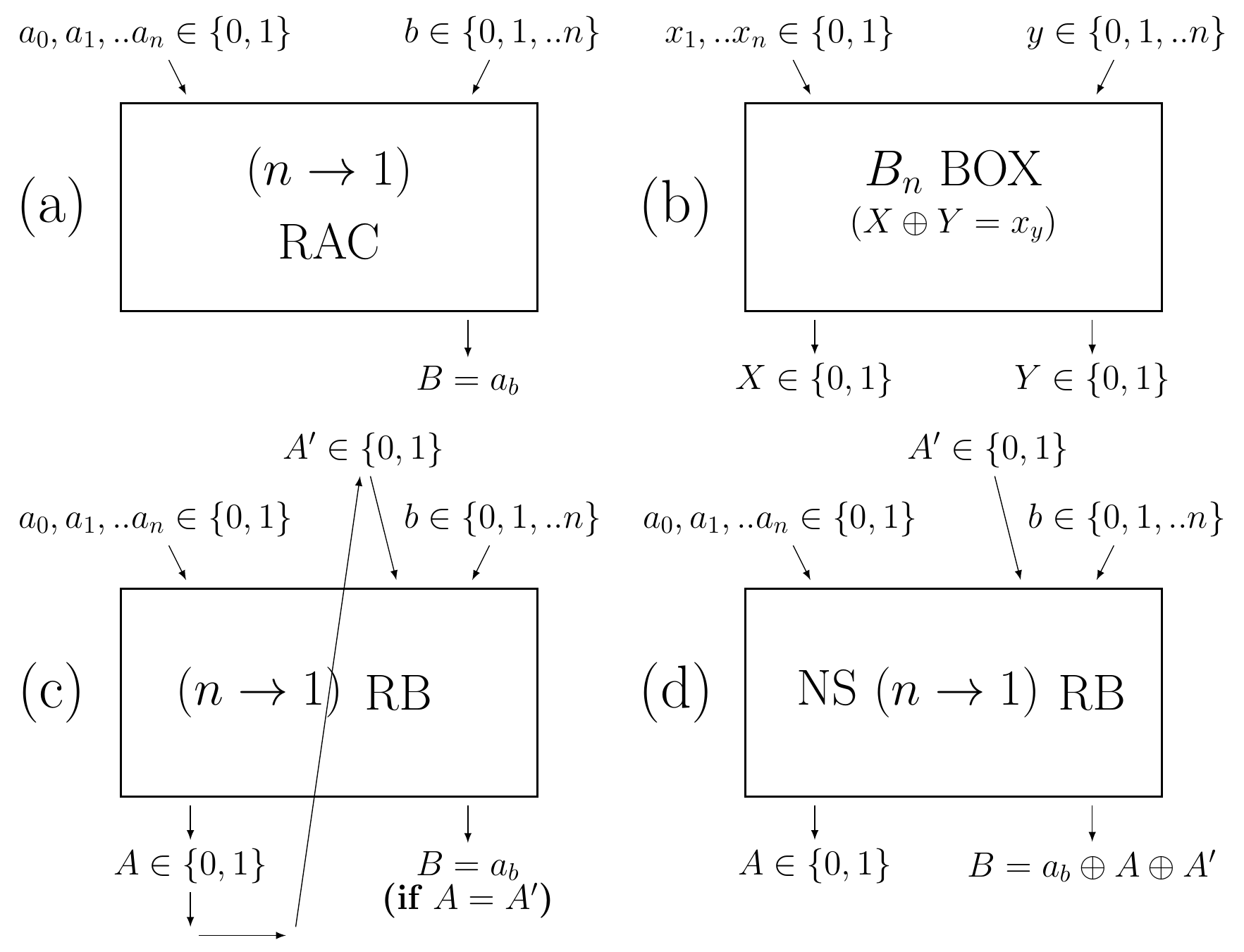}
\caption{  \label{fig1} (a) $(n\rightarrow 1)$ RAC. (b) $B_n$-BOX (c) $(n\rightarrow 1)$ RB acts like an $n\rightarrow 1$ RAC, provided that $A'=A$. In particular when $A$ is sent as a message to Bob and he inputs it into $A'$, then $B=a_b$. (d) No-signaling $n\rightarrow 1$ RB satisfies $B=a_b\oplus A \oplus A'$.}
\end{figure}

\noindent Let us define a {\bf \boldmath$(n\rightarrow1)$ RAC}. This is a \textbf{box} wherein, Alice is assigned $n$ input bits $a_0,a_1,..a_{n-1}$. Bob is assigned an input $b\in\{0,1,2,..n-1\}$ to decide which of Alice's bit he gets. Bob has a bit of output $B$. For a $(n\rightarrow1)$ RAC when $B=a_b$ for all possible inputs [see Fig. \ref{fig1} ]. \\
Consider a box that has an additional output on Alice's side $A$ and one more input $A'$ on Bob's side [see Fig. \ref{fig1} ]. Further suppose that it is no-signaling from Bob to Alice. Such a box we call $(n\rightarrow1)$ RB when the following condition holds: if $A=A'$, then it acts like a $(n\rightarrow1)$ RAC i.e., $B=a_b$. However when $A\neq A'$ we do not put any restrictions.  This box is described by a probability distribution $P(A,B|a_0,a_1,..a_{n-1},A',b)$ with a condition i.e., for all $i\in\{0,1,2,..n-1\}$,
\begin{equation}\label{e1}
P(B=a_i|A'=A,b=i)=1
\end{equation}

\noindent Notice that we have the freedom to define the probability distribution of $(n\rightarrow 1)$ RB as long as it can be turned into a perfect $(n\rightarrow 1)$ RAC. This implies we can have both signaling (only possible from Alice to Bob) and no-signaling  $(n\rightarrow 1)$ RB. \\ A \textit{no-signaling $(n\rightarrow1)$ RB} is a $(n\rightarrow1)$ RB with an additional condition, namely when no message is sent Bob should not able gain any information about Alice's inputs (no-signaling condition) i.e. $P(B|a_0,a_1,..a_{n-1},A',b)=P(B|a_0',a_1',..a_{n-1}',A',b)$ for all possible values of $B,A',b$. Now we characterize no-signaling $(n\rightarrow 1)$ RB by the following lemma.
\begin{mydef2} \label{lem1}
A no signaling $(n\rightarrow 1)$ RB for $A\neq A'$ acts as an anti-$(n\rightarrow 1)$ RAC, i.e., it satisfies
\begin{equation} \label{el1}
B=a_b\oplus A \oplus A'
\end{equation}
\end{mydef2}
Let Alice and Bob share a no-signaling $(n\rightarrow 1)$ RB. Suppose Alice \textit{does not} send the message and Bob chooses $A'$ randomly, 
\begin{equation}\label{e2}
P(A=A')=P(A\neq A')=\frac{1}{2}
\end{equation}
Let $P(B=a_i|b=i)$ denote the probability that Bob's outcome is correct when no message was sent. The \textit{no-signaling} condition along with the fact that Alice's inputs are uniformly distributed implies $P(B=a_i|b=i)=\frac{1}{2}$ [see Fig. 1]. Upon conditioning on the events $A=A'$ and $A\neq A'$ we obtain, 
\begin{eqnarray}
P(B=a_i|A=A',b=i)P(A=A)+\nonumber \\ P(B=a_i|A\neq A',b=i)P(A\neq A')=\frac{1}{2}
\end{eqnarray}
From (\ref{e1}) and (\ref{e2}), 
\begin{equation}
\frac{1}{2}+P(B=a_i|A\neq A',b=i)\frac{1}{2}=\frac{1}{2}
\end{equation}
implies, 
\begin{equation}
P(B=a_i|A\neq A',b=i)=0
\end{equation}
If $A\neq A'$, $B=a_i\oplus 1$ or, 
\begin{equation} \label{e7}
P(B=a_i\oplus 1|A\neq A',b=i)=1
\end{equation}
Thus equations (\ref{e1}) and (\ref{e7}) lead to desired result  (\ref{el1}). $\blacksquare$ \\
We shall now present an instance of a \textit{signaling $(n\rightarrow1)$ RB} which performs its duty regarding $(n\rightarrow 1)$ RAC when supplemented with a bit of communication but is signaling from Alice to Bob. It is a $(n\rightarrow1)$ RB  with an additional condition,
\begin{equation}
P(B=a_i|A\neq A',b=i)=\frac{1}{2}
\end{equation}
In this case when no message is sent Bob could still gain some information about Alice's inputs (no-signaling condition) as,
\begin{equation}
P(B=a_i|b=i)=\frac{3}{4}
\end{equation}
Finally we define our contender from non-local resources as a generalization to the PR. A  $B_n$-BOX is a bipartite \textit{no signaling} \textit{resource} (correlation) wherein, Alice's side has $n-1$ input bits $x_{1},x_{2},x_{3}...x_{n-1}$ and an output bit $X$. Bob's side has $n$ possible inputs corresponding to a input $n$it $y\in\{0,1,2,3...n-1\}$ and an output bit $Y$. A $B_n$-BOX is described by the probability distribution $P(X,Y|x_1,x_2,..x_{n-1},y)$ such that,
\begin{equation}
P(X,Y|x_1,x_2,..x_{n-1},y)=
\begin{cases}
\frac{1}{2} & for X\oplus Y=x_y,\\
0 & else.
\end{cases}
\end{equation}
The condition,
\begin{equation}
X\oplus Y=x_y
\end{equation}
will be called $B_n$ correlations, where $x_0=0$. 
\subsection*{Relationships}
\noindent We shall proof the following theorem which deals with equivalence between a $B_n$-BOX and a no-signaling $(n\rightarrow 1)$ RB.
\begin{mydef1} \label{thm1}
A $B_n$-BOX and a no-signaling $(n\rightarrow 1)$ RB are strictly equivalent.
\end{mydef1}
In order to prove equivalences between two resources it is necessary and sufficient to show that each can simulate the other. Here we provide a protocol using which Alice and Bob sharing a $B_n$-BOX can win a $(n\rightarrow1)$ RAC. 
\begin{enumerate}
\item Alice receives $n$ input bits $a_0,a_1,..a_{n-1}$ and she inputs $x_i=a_0\oplus a_i$ for $i\in\{1,2,..n-1\}$.
\item Alice obtains an output bit $X$ from $B_n$-BOX sends the message $m=a_0\oplus X$.
\item  Bob receives $y=b$ and obtains output bit  $Y$. 
\item Bob outputs the final answer $B=m\oplus Y=a_0\oplus X \oplus Y$.  
\end{enumerate}
\noindent Now for a $B_n$-BOX, $X\oplus Y=x_y$. Therefore $B=a_0\oplus x_y=a_i$ if $y\in\{1,2,..n-1\}$ and $B=a_0$ if y=0.

\noindent Now to complete the proof, we provide a protocol using which Alice and Bob sharing a no-signaling  $(n\rightarrow1)$ RB can simulate the statistics of $B_n$ BOX.
\begin{enumerate}
\item Alice receives $n-1$ input bits $x_1,x_2,..x_{n-1}$ and she inputs $a_i=x_i$ for $i\in\{1,2,..n-1\}$ and fixes $a_0=0$.
\item Alice obtains an output bit $X=A$ from the $(n\rightarrow 1)$ RB.
\item Bob receives $y\in \{0,1,2,..n-1\}$ and fixes $A'=0$ obtains output bit  $Y=B$.
\end{enumerate}
\noindent Observe whenever $X=A=A'=0$, using (\ref{e1}) we get $Y\oplus X=Y=B=a_b=x_i$. Further whenever $X=A\neq 0$, using (\ref{e7}) we obtain $Y\oplus X=Y\oplus 1=B\oplus 1=a_b\oplus 1 \oplus 1=a_b=x_i$. $\blacksquare$ \\ We shall now provide the following resource inequality which implies that having access to any $(n\rightarrow1)$ RB (signaling or no-signaling), one bit of communication (c-bit) and one shared random bit (sr-bit) we can simulate a $B_n$-BOX and additionally obtain erasure channel $\xi$ with probability of erasure $\epsilon=p(y\neq 0)$.
\begin{mydef3}\label{re1}
between a $(n\rightarrow1)$ RB and a $B_n$-BOX : \\
We show that the following inequality holds for any $(n\rightarrow1)$ RB,
\begin{eqnarray} 
(n\rightarrow1) RB  +1 c-bit + 1  sr-bit \geq \nonumber \\ B_n-BOX + \xi
\end{eqnarray}
where $\xi$ is a bit erasure channel.
\end{mydef3}
 Since by definition $(n\rightarrow1)$ RB plus 1 bit of communication offers a RAC we shall prove the following inequality instead,
\begin{eqnarray}
(n\rightarrow1) RAC + 1  sr-bit \geq \nonumber \\ B_n-BOX + \xi
\end{eqnarray}
In order to reproduce a $B_n$-BOX in the case when $y=0$, one can use just shared randomness since Alice's and Bob's outputs must be the same i.e., $X\oplus Y=0$. The $(n\rightarrow1)$ RAC is not used up and can be utilized for communication of the bit $a_0$. But when $y\neq 0$, Bob will need the $(n\rightarrow1)$ RAC to reproduce $B_n$ correlations as $X\oplus Y=x_y$, and in this case no communication will be performed. \\
Let $z$ denote the bit to be communicated. Alice puts $a_0=z$ and  $a_i=x_i$ where $i\in \{1,2,..n-1\}$, while Bob inputs $b=y$. Alice and Bob use shared random bits for outputs. When $y=0$, Bob simply outputs the shared random bit and $B_n$-BOX is reproduced. When $y\neq 0$, Bob performs a CNOT gate with his output $B$ being the control bit and his shared random bit being target bit. When $y\neq 0$ we need to have correlations when $x_i=0$ and anti correlations when $x_i=1$ given that Bob inputs $b=y=i$. From the definition of a $(n\rightarrow 1)$ RAC, when $b=y=i$ where $i\in \{1,2,..n-1\}$, we have $B=x_i$. Hence, when $x_i=0$ the shared random bit is bot flipped and Alice and Bob have correlations and when $x_i=1$ the bit is flipped and they have anti correlations. Thus the protocol perfectly simulates a $B_n$-BOX.\\
When $y=0$, Bob's output $B=a_0=z$, hence the message is perfectly transmitted, whereas for $y\neq 0$ Bob's output $B=x_y$ and the message is lost. Thus we obtain a erasure channel with probability of erasure  $\epsilon=P(y\neq 0)=\frac{n-1}{n}$ (assuming Bob's input are uniformly distributed). \\
\textit{Tightness of resource inequality \ref{re1}:} This resource inequality is trivial for a no-signaling $(n\rightarrow 1)$ RB. However using the signaling $(n\rightarrow 1)$ RB defined above we can tighten the inequality through the following theorem.
\begin{mydef1} \label{thm2}
Assume that $x_1,x_2,..x_n,y$ are generated uniformly at random. Let us suppose for the signaling $(n\rightarrow 1)$ RB described above, a channel $\Lambda$ satisfies the following inequality:
\begin{eqnarray}
(n\rightarrow1) RB + 1  c-bit \geq \nonumber \\ B_n-BOX + \Lambda
\end{eqnarray}
Then the capacity of bit channel $\Lambda$ is upper bounded by $\frac{1}{n}$.
\end{mydef1} \label{thm2}
For the proof see Appendix A. The theorem shows that in order to simulate a $B_n$-BOX by such a no-signaling $(n\rightarrow 1)$ RB, we need, in addition at-least $\frac{n-1}{n}$ bit of communication. Thus, in this respect the signaling $(n\rightarrow 1)$ RB is  weaker than a no-signaling $(n\rightarrow 1)$ RB.
\subsection*{$(2\rightarrow 1)$ RACs as building blocks for $(n\rightarrow1)$ RAC}
\begin{figure} 
\includegraphics[scale=0.5]{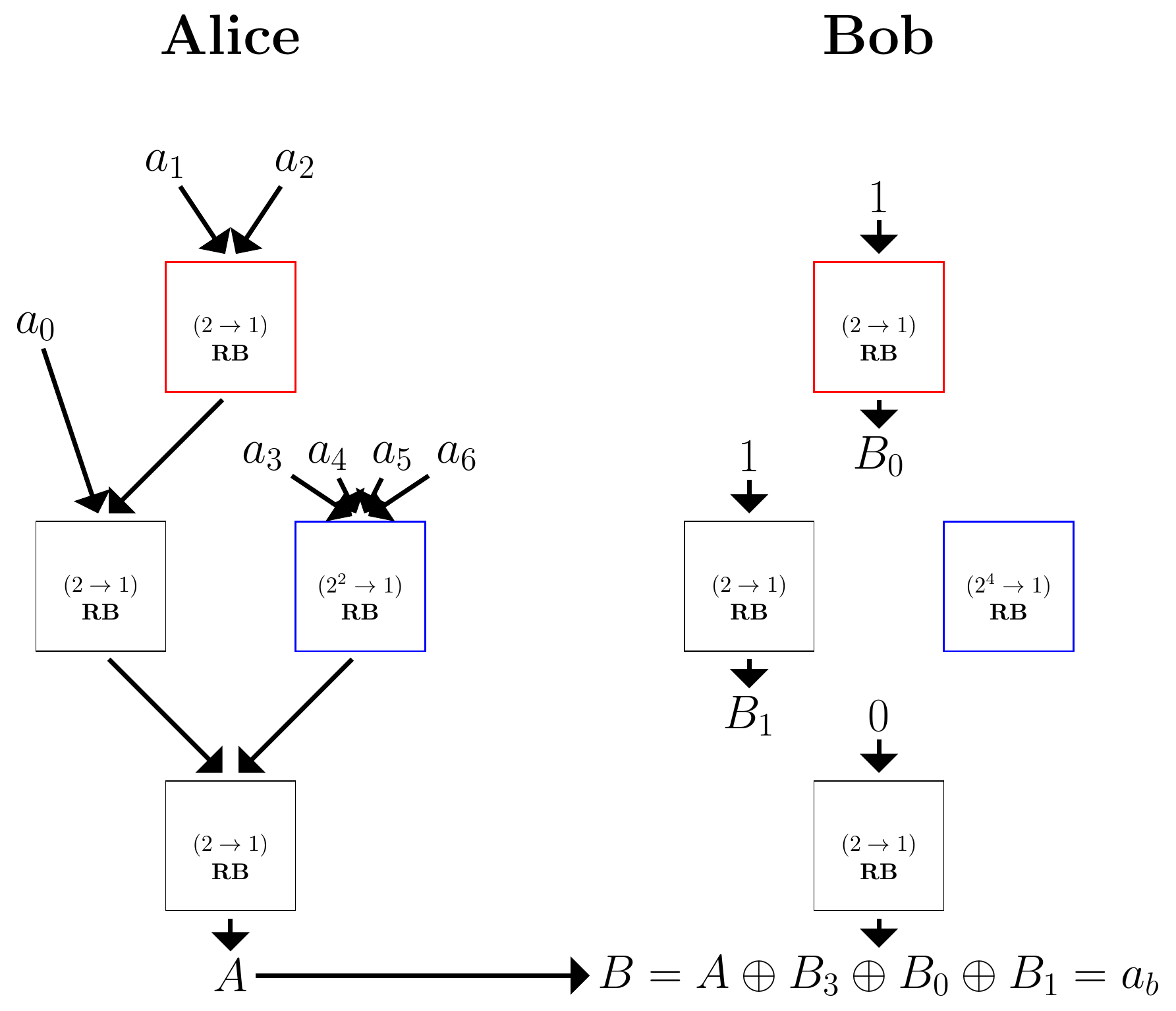}
\caption{ \label{7to1} (Color online) The figure demonstrates the use of CONCATENATION and ADDITION of no-signaling $(2\rightarrow 1)$
RBs and 1 c-bit
to win a $(7\rightarrow 1)$ RAC  . \textit{RESOURCE: $7-1=6$ } no-signaling $(2\rightarrow 1)$
RB. In particular, here Bob is trying to learn $a_2$.}
\end{figure}

\noindent It is a known fact that some number of no-signaling $(2\rightarrow 1)$ RBs and 1 c-bit can be used to construct a general  $(n\rightarrow 1)$ RAC. Through the following theorem we shall show that the minimum number of no-signaling $(2\rightarrow 1)$ RBs to win a  $(n\rightarrow 1)$ RAC when Alice is allowed to communicate 1 c-bit is $n-1$.
\begin{mydef1} \label{thm3}
$n-1$ no-signaling $(2\rightarrow 1)$ RBs  are necessary and sufficient to win a $(n\rightarrow 1)$ RAC when Alice is allowed to communicate 1 c-bit.
\end{mydef1} 
We shall prove this theorem in two parts. First we shall prove the following lemma,
\begin{mydef2} \label{lem:countRAC}
$n-2$ no-signaling $(2\rightarrow 1)$ RBs + 1 c-bit cannot win a  $(n\rightarrow 1)$ RAC.
\end{mydef2}
For the proof see Appendix B. To complete the proof we provide a protocol which uses $n-1$ no-signaling $(2\rightarrow 1)$ RB and 1 c-bit of additional communication to win a $(n\rightarrow 1)$ RAC. The protocol we provide uses two subroutines namely, concatenation and addition (see Appendix C). For instance a $(7\rightarrow1)$ RAC requires 6  no-signaling $(2\rightarrow 1)$
RBs and a c-bit of communication [see Fig. \ref{7to1} ]

\section{From bits to dits.}
In this section we provide further generalization based on the dimension $d$ of inputs provided to Alice and message she sends to $(n\rightarrow1)$ RAC and $B_n$-BOX. We start of with defining the resources under consideration and subsequently study their relationships.
\subsection*{$(n\rightarrow1,d)$ RAC, $(n\rightarrow1,d)$ RB, $B_n^d(+)$-BOX and $B_n^d(-)$-BOX}
\begin{figure}
\includegraphics[scale=0.5]{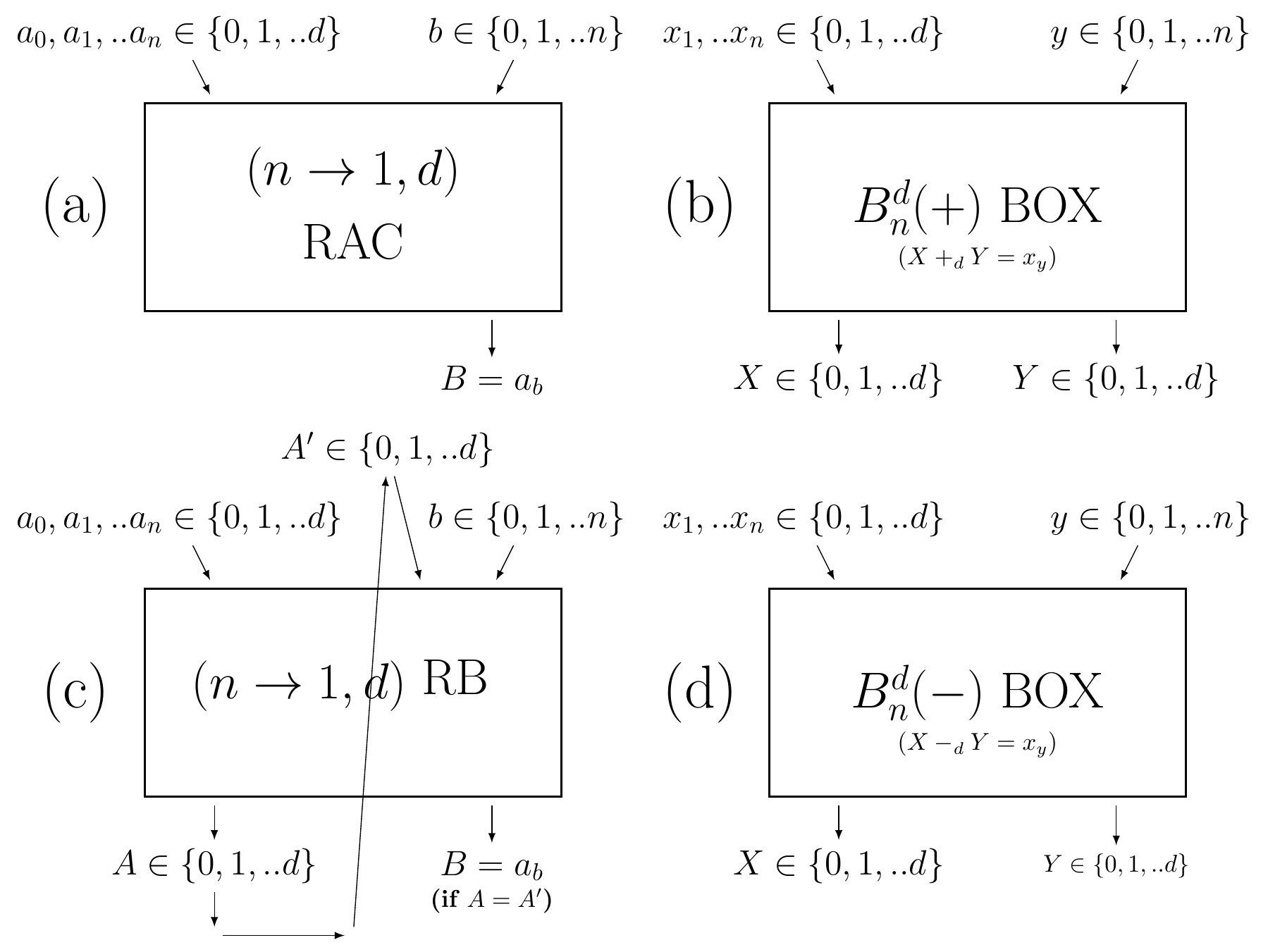}
\caption{  \label{Def2} (a) $(n\rightarrow 1,d)$ RAC. (b) $B_n^d(+)$-BOX (c) $(n\rightarrow 1,d)$ RB acts like an $(n\rightarrow 1)$ RAC, provided that $A'=A$. In particular when $A$ is sent as a message to Bob and he inputs it into $A'$, then $B=a_b$. (d) $B_n^d(-)$-BOX.}
\end{figure}

\noindent We start by defining a $(n\rightarrow1,d)$ RAC. This is a box wherein,  Alice is assigned $n$ input $d$its $a_0,a_1,..a_{n-1}$ where $a_i\in \{0,1,..d-1\}$ for $i\in \{0,1,..n-1\}$.  Bob is assigned an input $b\in\{0,1,2,..n-1\}$ to decide which of Alice's $d$it he gets. Bob has a $d$it of output $B$. Such a box is a $(n\rightarrow1,d)$ RAC when $B=a_b$ for all possible inputs [see Fig. \ref{Def2}]. \\
Let us consider another box that has an additional output $d$it on Alice's side $A$ and one more input $d$it $A'$ on Bob's side [see Fig. \ref{Def2}]. Further suppose that it is no-signaling from Bob to Alice. Such a box we call $(n\rightarrow1,d)$ RB when the following condition holds: if $A=A'$, then it acts like a $(n\rightarrow1,d)$ RAC i.e., $B=a_b$. However when $A\neq A'$ we do not put any restrictions.  This box is described by a probability distribution $P(A,B|a_0,a_1,..a_{n-1},A',b)$ with a condition i.e., for all $i\in\{0,1,2,..n-1\}$,
\begin{equation}\label{de1}
P(B=a_i|A'=A,b=i)=1
\end{equation}
This box is designed such that when supplemented with one bit of communication, it offers a perfect $(n\rightarrow1,d)$ RAC.
A \textit{no-signaling $(n\rightarrow1,d)$ RB} is a $(n\rightarrow1,d)$ RB with an additional condition, namely when no message is sent Bob should not able gain any information about Alice's inputs (no-signaling condition), i.e. $P(B|a_0,a_1,..a_{n-1},A',b)=P(B|a_0',a_1',..a_{n-1}',A',b)$ for all possible values of $B,A',b$ [see Fig. 3]. \\
The case of $d>2$ presents itself with intricate details. To see this let Alice and Bob share a no-signaling $(n\rightarrow 1,d)$ RB. Suppose Alice \textit{does not} send the message and Bob chooses $A'$ randomly, 
\begin{equation}\label{de}
P(A=A')=\frac{1}{d}
\end{equation}
Let $P(B=a_i|b=i)$ denote the probability that Bob's outcome is correct when no message was sent. The \textit{no-signaling} condition along with the fact that Alice's inputs are uniformly distributed implies $P(B=a_i|b=i)=\frac{1}{d}$. Now, 
\begin{eqnarray}
P(B=a_i|b=i)=P(B=a_i,A=A'|b=i)+\nonumber \\ P(B=a_i,A\neq A'|b=i)=\frac{1}{d}
\end{eqnarray}
\begin{eqnarray}
P(B=a_i|b=i)=P(B=a_i|A=A',b=i)P(A=A)+\nonumber \\ P(B=a_i|A\neq A',b=i)P(A\neq A')=\frac{1}{d}
\end{eqnarray}
From (\ref{de1},\ref{de}), 
\begin{equation}
\frac{1}{d}+P(B=a_i|A\neq A',b=i)\frac{d-1}{d}=\frac{1}{d}
\end{equation}
implies, 
\begin{equation}\label{imp}
P(B=a_i|A\neq A',b=i)=0
\end{equation}
Notice that this \textit{does not} completely specify the value(or probability distribution) of $B$ except for the fact that it must not be $a_i$, as compared to the case of $d=2$. Thus, even under no-signaling for the case $d>2$ we have the freedom to define probability distribution of no-signaling $(n\rightarrow 1,d)$ RB as long as it can be turned into a perfect $(n\rightarrow 1,d)$ RAC. This implies we can define subclasses of no-signaling $(n\rightarrow 1,d)$ RB based on additional condition over the probability distribution. W.l.o.g when no message is sent we assume that Bob always inputs $A'$, then we have the following three different no-signaling $(n\rightarrow 1,d)$ RB,
\begin{itemize}
\item \textit{no-signaling $(n\rightarrow 1,d)$ RB (+)} : This particular instance of no-signaling $(n\rightarrow1,d)$ RB is defined by additional condition over probability distribution $P(B+_d A=a_i|A,A'=0,b=i)=1$  (where $+_d$  is addition modulo $d$).
\item \textit{no-signaling $(n\rightarrow 1,d)$ RB (-)} : This one is defined by additional condition over probability distribution $P(B-_d A=a_i|A,A'=0,b=i)=1$  (where $-_d$  is subtraction modulo $d$).
\item \textit{no-signaling $(n\rightarrow 1,d)$ RB (3)} :This particular instance of no-signaling $(n\rightarrow,d)$ RB is defined by additional condition over probability distribution $P(B=j|A\neq0,A'=0,b=i)=\frac{1}{d}$ for $j\in\{0,1,2,..d-1\}-{a_i}$. This is a bad instance in the sense that while it fulfills its duties as a $(n\rightarrow 1,d)$ RAC when $A'=A$ but it cannot simulate either $B_n^d(+)$-BOX or $B_n^d(-)$-BOX. 
\end{itemize}
The case of $d>2$ is also rich in complexity when it comes to defining a generalization to the PR (or a $B_n$-BOX). For instance we define two possible generalizations namely, $B_n^d(+)$-BOX and $B_n^d(-)$-BOX. A $B_n^d(+)$-BOX is a bipartite \textit{no signaling} \textit{resource} (correlation) wherein, Alice's side has $n-1$ input $d$its $x_{1},x_{2},x_{3}...x_{n-1}$ and an output $d$it $X$. Bob's BOX has $n$ possible inputs corresponding to a input $n$it $y\in\{0,1,2,3...n-1\}$, and receives output $d$it $Y$. 
A $B_n^d(+)$-BOX is described by the probability distribution $P(X,Y|x_1,x_2,..x_{n-1},y)$ such that, \\
\begin{equation}
P(X,Y|x_1,x_2,..x_{n-1},y)=
\begin{cases}
\frac{1}{d} & for X+_d Y=x_y,\\
0 & else.
\end{cases}
\end{equation}
The condition,
\begin{equation}
X+_d Y=x_y
\end{equation}
will be called $B_n^d(+)$ correlations.\\
Similarly a $B_n^d(-)$-BOX is described is described by the probability distribution $P(X,Y|x_1,x_2,..x_{n-1},y)$ such that, \\
\begin{equation}
P(X,Y|x_1,x_2,..x_{n-1},y)=
\begin{cases}
\frac{1}{d} & for X-_d Y=x_y,\\
0 & else.
\end{cases}
\end{equation}
The condition,
\begin{equation}
X-_d Y=x_y
\end{equation}
will be called $B_n^d(-)$ correlations.\\
\subsection*{Relationships}
\noindent We begin with showing through the following theorem that shows $B_n^d(+)$-BOX is equivalent to a no-signaling $(n\rightarrow 1,d)$ RB (+).
\begin{mydef1} \label{thm4}
A $B_n^d(+)$-BOX and a no-signaling $(n\rightarrow1,d)$ RB (+) are equivalent for  $d>2$.
\end{mydef1}
We will prove the above theorem by giving explicit protocols.
We provide a protocol using which Alice and Bob sharing a $B_n^d(+)$-BOX and 1 c-$d$it of communication can win a $(n\rightarrow1,d)$ RAC with certainty. 
\begin{enumerate}
\item Alice receives $n$ input $d$its $a_0,a_1,..a_{n-1}$ and she inputs $x_i=a_i-_d a_0$ for $i\in\{1,2,..n-1\}$.
\item Alice obtains an output $d$it $X$ from $B_n^d(+)$ BOX sends the message $m=X+_d a_0$.
\item Bob receives $y\in\{0,1,2,..n-1\}$ and obtains output $d$it  $Y$. 
\item Bob outputs the final answer $B=m+_dY=a_0+_d X +_d Y$.
\end{enumerate}
\noindent Now for a $B_n^d(+)$-BOX, $X+_d Y=x_y$. Therefore $B=a_0+_dx_y=a_i$ if $y\in\{1,2,..n-1\}$ and $B=a_0$ if y=0. \\
Finally to complete to proof we provide a protocol using which Alice and Bob sharing no-signaling $(n\rightarrow1,d)$ RB (+) can simulate a $B_n^d(+)$-BOX perfectly.
\begin{enumerate}
\item Alice receives $n-1$ input $d$its $x_1,x_2,..x_{n-1}$ and she inputs $a_i=x_i$ for $i\in\{1,2,..n-1\}$ and fixes $a_0=0$.
\item Alice obtains an output bit $X=A$ from the no-signaling $(n\rightarrow 1,d)$ RB (+).
\item Bob receives $y\in \{0,1,2,..n-1\}$ and fixes $A'=0$ obtains output $d$it  $Y=B$.
\end{enumerate}
\noindent  Observe whenever $X=A=A'=0$, $Y+_d X=Y=B=a_b=x_i$. Further whenever $X=A\neq 0$, $Y+_dX=B+_dA=a_b=x_i$. $\blacksquare$ \\
\begin{mydef1} \label{thm5}
A $B_n^d(-)$-BOX and a no-signaling $(n\rightarrow1,d)$ RB (-) are strictly equivalent for  $d>2$.
\end{mydef1}
The proof is omitted as it is similar to that of Theorem \ref{thm4}. $\blacksquare$
\\ We shall now provide the following resource inequality which implies that having access to any $(n\rightarrow1,d)$ RB, one $d$it of communication (c-$d$it) and one shared random $d$it (sr-$d$it) we can simulate a $B_n^d(+)$-BOX (or $B_n^d(-)$-BOX) and additionally obtain erasure $d$it channel $\xi_d$ with probability of erasure $\epsilon=p(y\neq 0)$:
\begin{mydef3}\label{re2}
between a $(n\rightarrow1,d)$ RB and a $B_n^d(+)$-BOX : \\
We show that the following inequality holds for any $(n\rightarrow1,d)$ RB,
\begin{eqnarray} 
(n\rightarrow1,d) RB  +1 c-dit + 1  sr-dit \geq \nonumber \\ B_n^d(+)-BOX + \xi_d
\end{eqnarray}
where $\xi_d$ is a $d$it erasure channel. 
\end{mydef3}
Since by definition $(n\rightarrow1,d)$ RB plus 1 $d$it of communication offers a $(n\rightarrow1,d)$ RAC we shall prove the following inequality instead ,
\begin{eqnarray}
(n\rightarrow1,d) RAC + 1  sr-dit \geq \nonumber \\ B_n^d(+)-BOX + \xi_d
\end{eqnarray}
In order to reproduce a $B_n^d(+)$-BOX (or $B_n^d(-)$-BOX) in the case when $y=0$, one can use just shared randomness of the form $s_A +_d s_B =0$. The $(n\rightarrow1,d)$ RAC is not used up and can be utilized for communication of the $d$it $a_0$. But when $y\neq 0$, Bob will need the $(n\rightarrow1,d)$ RAC to reproduce $B_n^d(+)$ correlations as $X+_d Y=x_y$, and in this case no communication will be performed. \\
Let $z$ denote the $d$it to be communicated. Alice puts $a_0=z$ and  $a_i=x_i$ where $i\in \{1,2,..n-1\}$, while Bob inputs $b=y$. Alice and Bob use shared random $d$its for outputs. When $y=0$, Bob simply outputs the random $d$it and $B_n^d(+)$-BOX is reproduced. When $y\neq 0$, Bob adds $B$ to her shared random bit $s_B$. From the definition of a $(n\rightarrow 1,d)$ RAC, when $b=y=i$ where $i\in \{1,2,..n-1\}$, we have $B=x_i$. Hence, when $x_i=0$ the shared random bit remains the same so that $X+_d Y=s_A+_d s_B=0$ and when $x_i\neq 0$ Bob produces $Y=s_B+_d B=s_B+_d x_i$ so that $X+_d Y=s_A+_ds_B+_d x_i=x_i$ and in this case the message is lost. Thus we obtain a $d$it erasure channel with probability of erasure  $\epsilon=P(y\neq 0)=\frac{n-1}{n}$ (assuming Bob's input are uniformly distributed). \\
Now we proceed to show that the no-signaling $(n\rightarrow1,d)$ RB (3) cannot simulate the $B^d_n(+)$-BOX. We shall use the case of $n=2,d=3$ for simplicity.\\
\textit{Tightness of resource inequality \ref{re2}:} This resource inequality is trivial for a no-signaling $(2\rightarrow 1,3)$ RB (+) when trying to simulate $B^3_2(+)$-BOX or no-signaling $(2\rightarrow 1,3)$ RB (-) trying to simulate $B^3_2(-)$-BOX  . However using the no-signaling $(2\rightarrow 1,3)$ RB (3) defined above we can tighten the inequality through the following theorem.
\begin{mydef1} \label{thm6}
Assume that $x,y$ (inputs to the $ B_2^3(+)$-BOX)  are generated uniformly at random. Let us suppose for the no-signaling $(2\rightarrow 1,3)$ RB (3) described above, a channel $\lambda_3$ satisfies the following inequality:
\begin{eqnarray}
(2\rightarrow1,3) RB + 1  c-3it \geq \nonumber \\ B_2^3(+)-BOX + \Lambda_3
\end{eqnarray}
Then the mutual information of $3$it channel is upper bounded by $\frac{1}{2}$.
\end{mydef1}
For the proof see Appendix D. The theorem shows that in order to simulate a $B_2^3(+)$-BOX by such a no-signaling $(2\rightarrow 1,3)$ RB (3), we need, in addition at-least $\frac{1}{2}$ $3$it of communication. Thus, in this respect the no-signaling $(2\rightarrow 1,3)$ RB (3) is  weaker than a no-signaling $(2\rightarrow 1,3)$ RB (+) or no-signaling $(2\rightarrow 1,3)$ RB (-). It is straightforward to generalize above theorem for arbitrary $n,d$. 
\begin{mydef1} \label{thm7}
Assume that $x_1,x_2,..x_n,y$  are generated uniformly at random. Let us suppose for the no-signaling $(n\rightarrow 1,d)$ RB (3) described above, a channel $\lambda_d$ satisfies the following inequality:
\begin{eqnarray}
(n\rightarrow1,d) RB + 1  c-dit \geq \nonumber \\ B_d^n(+)-BOX + \Lambda_d
\end{eqnarray}
Then the mutual information of $d$it channel is upper bounded by $\frac{1}{n}$.
\end{mydef1}
The proof of the above theorem follows directly from the proof of Theorem 2 and \ref{thm6}.
\section{Conclusions and Further Directions}
\noindent In this work we introduced generalizations of a static no-signaling non-local resource PR based on number of inputs provided to Alice, namely $B_n$-BOX and then based on the dimension of the inputs provided to Alice, namely $B_n^d(+)$-BOX and $B_n^d(-)$-BOX. \\
In the former case we show that a $B_n$-BOX can win with a certainty a functionality $(n\rightarrow 1)$ RAC. Furthermore a no-signaling $(n\rightarrow 1)$ RB can simulate a $B_n$-BOX. Hence the two resource are shown to be equivalent. We bring up a signaling $(n\rightarrow 1)$ RB and show that it cannot simulate the $B_n$-correlation. To quantify the above we provide a resource inequality and show it is saturated. As an application to the above we prove that under the restriction that Alice is only allowed to communicate 1 cbit of communication we require at-least $(n-1)$ no-signaling $(2\rightarrow 1)$ RB (or PR) in order to win a $(n\rightarrow 1)$ RAC. \\
In the latter case of dimension $d>2$ we find that the no-signaling condition is not enough to enforce a strict equivalence between $B_n^d(+)$-BOX (or $B_n^d(-)$-BOX) and no-signaling $(n\rightarrow1,d)$ RB. We introduce three classes of no-signaling $(n\rightarrow1,d)$ RB, namely $(+)$,$(-)$,$(3)$. We show that a $B_n^d(+)$-BOX ($B_n^d(-)$-BOX is strictly equivalent to no-signaling $(n\rightarrow1,d)$ RB $(+)$ (or $(-)$). However $(n\rightarrow1,d)$ RB $(3)$ cannot simulate $B_n^d(+)$-BOX (or $B_n^d(-)$-BOX). Finally to quantify the same we provide a resource inequality and show that it is saturated.\\
We have shown that in the case of higher dimension $d>2$ there exists no-signaling $(n\rightarrow1,d)$ RB which cannot simulate all of extremal non-local resources which can be used to win a $(n\rightarrow1,d)$ RAC. However the question remains open that whether there exists any extremal non-local resource which cannot win a $(n\rightarrow1,d)$ RAC. \\
Such equivalences can be generalized to multiparty scenario. In \cite{A70747}  we study equivalence between $n$-party Svetlichny BOXes and a generalization of the standard two party RAC , Controlled Random Access Codes (C-RAC).
\section{Acknowledgments}
We acknowledge useful discussions with Micha{\l} Horodecki.
This project was supported by grant, Harmonia 4 (Grant number: UMO-2013/08/M/ST2/00626), ERC AdG QOLAPS. AC would like to acknowledge CCNSB/CSTAR of IIIT-Hyderabad and  support by Prof. Indranil Chakrabarty. KH acknowledges grant Sonata Bis 5 (Grant number: 2015/18/E/ST2/00327).
\bibliography{sample.bib}

\section*{Appendix A: proof of theorem 2}
We provide the proof for resource inequality \ref{re1} and consequently the in-equivalence of signaling $(n\rightarrow 1)$ RB and $B_n$-BOX. We give the proof in two parts: 
\begin{enumerate}
\item We shall show that if the bit of communication is not used to send the output of Alice's RB $A$ then $B_n$-BOX cannot be obtained..
\item If the bit of communication is used to send $A$ and $B_n$ BOX is obtained, then the capacity of obtainable channel $\Lambda$ is upper bounded by $\frac{1}{n}$ bit. 
\end{enumerate}
\subsection*{Part I}
The part I says that if we do not input $A$ to $A'$ then $B_n$-BOX cannot be obtained.\\
Let us denote $m$ for the one-bit message to be communicated to Bob. The goal is to obtain perfect $B_n$ correlations i.e. $Y=X\oplus x_y$ in any case $m=0$ or $1$. Bob's output for any give $m$ in depends on RAC's settings on Bob's side: $Y=Y(b,A',B)$. For any fixed $m=m_0$ there are two possible cases $A'=A$ and $A\neq A'$. In the first case $B_n$ correlations are obtained using by processing a perfect RAC. But in the case $A'\neq A$ the signaling $n \rightarrow 1$ RB offers a random $B$ which does not depend on the work of RAC. Hence $Y$ can be obtained solely from processing of $y$ i.e. $Y=Y(y)$. Since we want to obtain perfect $B_n$ correlations $Y(y=0)=X$ and $Y(y\neq 0)=X\oplus x_y$. By adding  $Y(y=0)$ and $Y(y\neq 0)$ Bob can compute $x_y$. We therefore obtain, that in the case $A'\neq A$, the value of $x_y$ must be know to Bob.\\
However signaling $n\rightarrow 1$ RB is no-signaling from Bob to Alice. For $y\in \{0,1,..n-1\}$ and $n>2$ Alice cannot know in advance the value of $y$ in order to send $m=x_y$. These leaves only one option to send $A$ which shall be dealt with in the next part $\blacksquare$.
\subsection*{Part II}
We will show using information theoretic tools, that if the signaling $(n\rightarrow 1)$ RB considered in Theorem 2 supplemented with one bit of communication is to reproduce exactly $B_n$-BOX and some channel, then the mutual information of the channel must be bounded by $\frac{1}{n}$ (assuming that Alice's output of the RB $A$ will be inserted directly into as Bob's second input to the RB i.e. $A'=A$). \\
\textit{Assumptions:} Alice is given variables $x_1,x_2,..x_{n-1}$ and z. Bob is given variable $y\in \{0,1,..n-1\}$. Both are given access to common variable s such that $x_1,x_2,..x_{n-1},z,y,s$ are mutually independent. Alice generates $A$ from $x_1,x_2,..x_{n-1},z,s$ and inputs $a_0,a_1,..a_{n-1}$ to RAC. Bob generates $b$ from $y,s$ and inputs it into RAC. These strategies result in shared joint probability distribution $P(x_1,x_2,..x_{n-1},z,y,s,b,A',A,B)$, where $B=a_b$ is obtained from $(n\rightarrow 1)$ RAC on Bob's side, and $Y$ is generated out of $b,B,s,y$ by Bob. 
First we shall express the Theorem 2 in other words. Under Assumptions 1, if variables $x_1,x_2,..x_{n-1},y,A,B$ perfectly reproduce $B_n$ correlations, there holds:
\begin{equation}
I(z:B,b,y,s)\leq \frac{1}{n}
\end{equation}
where $z$ is the message that Alice sends to Bobs.
We shall prove the above in two parts:
\begin{enumerate}
\item First we shall use entropies and correlation to state the fact that to simulate the $B_n$ BOX Bob has to guess perfectly $X$ when $y=0$ and $x_y\oplus X$ when $y\neq 0$. 
\item Second we shall show that it is impossible to send more than 1 $3$it through a channel with 1 $3$it capacity. As in our case Alice would like to send both $x_1$ and $z$ which bounds Bob's possible information gain about $z$. 
\end{enumerate}
\begin{mydef2}
Under Assumptions 1, if variables $(x_1,x_2,x_3..x_{n-1},y,A,B)$ simulate perfectly $B_n$ correlations, there holds:

\begin{equation}
I(B:X|b,s,y= 0)=H(X|b,s,y=0)
\end{equation}
\begin{equation}
I(B:X\oplus x_y|b,s,y\neq 0)=H(X\oplus x_y|b,s,y\neq0)
\end{equation}
\end{mydef2}
In order to reproduce $B_n$-correlations given $y=0$, Bob should perfectly guess $X$, whereas given $y\neq 0$ he should perfectly guess $X\oplus x_y$. 
This implies that there must be $\max_j[p(a=j|B=l,b=k,y=0,s=i)]=1$. Then for $y=0$ the values of variables $B,b,s$ should determine uniquely the value of $X$ i.e. $H(X|B,b,s,y=0)=0$. In such a case, i.e. $I(X:B|b,s,y=0)=H(X|b,s,y=0)$. Analogously, we obtain $I(X\oplus x_n:B|b,s,y\neq 0)=H(X\oplus x_y|b,s,y\neq0)$. $\blacksquare$ \\
\textit{One cannot send more than one bit through a single-bit wire.} \\
Here, we prove the following theorem which provides the key argument in the proof of Theorem 2. Namely its shows a tradeoff between Bob's correlations with $X$ and $X\oplus x_y$ (that should be high if he simulates $B_n$ correlations) and his correlations with $z$.
\begin{mydef1}
Under aforementioned assumptions, there holds:
\begin{multline}
 \frac{1}{n}[\sum_{i=1}^{n-1}I(X\oplus x_i:B|b,s,y=i) + I(X:B|b,s,y=0)] +\\ 
I(z:B|b,s,y) \leq \\ 
\frac{1}{n}I(X:X\oplus x_1:X\oplus x_2:..:X\oplus x_{n-1}:z |b,s) 
+H(B|b,s,y)
\end{multline}
\end{mydef1}
In the proof for the above theorem we use the following fact, which captures that one cannot send reliably $n$ bits ($n>1$) through a single bit wire, unless the bits are correlated.
\begin{mydef2}
For any random variables $S_1,S_2,,..S_n,T,V$ there holds:
\begin{equation}
\sum_{i=1}^n I(S_i:T|V) \leq I(S_1:S_2:..S_n:T|V)
\end{equation}
where $I(S_1:S_2:..S_n:T|V)=\sum_{i=0}^{n}H(S_i|V)+H(T|V)-H(S_1,S_2,..S_n,T|V)$
\end{mydef2}

First we proof the above fact without conditioning. We shall use the following fact recursively. For any random variables $S_i,S_j,T$ it follows directly from strong subadditivity: 
\begin{equation}\label{e33}
H(S_iS_jU)+H(T)\leq H(S_iT) + H(S_jT)
\end{equation}
By expressing mutual information via Shannon entropies, we can rewrite LHS as:
\begin{multline}
nH(T)+\sum_{i=1}^{n}H(S_i)-\sum_{i=1}^{n}H(S_iT)
\end{multline}
Using (\ref{e33}) n-1 times we can upper bound LHS by,
\begin{equation}
H(T)+\sum_{i=1}^{n}H(S_i)-H(S_1,S_2,..S_n,T)\equiv I(S_1:S_2:..S_n:T)
\end{equation}
Which is the desired result without conditioning on $V$. We can now fix $V=v$, and the thesis will hold for conditional distribution $p(S_1,S_2,..S_n,T|V=v)$:
\begin{equation}
\sum_{i=1}^n I(S_i:T|V=v) \leq I(S_1:S_2:..S_n:T|V=v)
\end{equation}
The thesis is obtained after multiplying each side by $p(V=v)$, and summing over range of variable $V$. $\blacksquare$ \\
Moving on with the proof of Theorem 8. Let us reformulate LHS and fix s=j:
\begin{multline}
 \frac{1}{n}[\sum_{i=1}^{n-1}I(X\oplus x_i:B|b,s=j,y=i) + I(X:B|b,s=j,y=0)] \\ \nonumber
I(z:B|b,s=j,y) 
\end{multline}
By decomposing the last term into $n$ terms, which depend on the value of $y$ we obtain:
\begin{multline}
 \frac{1}{n}[\sum_{i=1}^{n-1}(I(X\oplus x_i:B|b,s=j,y=i) + I(z:B|b,s=j,y=i))+ \\
I(X:B|b,s=j,y=0) + I(z:B|b,s=j,y=0)]
\end{multline}
We use Lemma 4 (for $n=2$) pairwise to show that the above quantity is upper bounded by
\begin{multline}\label{e38}
\frac{1}{n}[\sum_{i=1}^{n-1}(I(X\oplus x_i:z|b,s=j,y=i) + \\ I(B:X\oplus x_i,z|b,s=j,y=i))+
I(X:z|b,s=j,y=0) + \\ I(B:X,z|b,s=j,y=0)]
\end{multline}
Observe that $(X\oplus x_i,z|s=j)$ is independent from $(y,b|s=j)$, hence there is $I(X\oplus x_i:z|b,s=j,y=i)=I(X\oplus x_i:z|b,s=j)$ and similarly $I(X:z|b,s=j,y=0)=I(X:z|b,s=j)$. Multiplying both sides these equalities by $p(s=i)$ and summing up over values of $s$ we get $I(X\oplus x_i:z|b,s,y=i)=I(X\oplus x_i:z|b,s)$ and $I(X:z|b,s,y=0)=I(X:z|b,s)$. Applying the same to (\ref{e38}) and using the latter equalities we obtain:
\begin{multline}\label{e38}
\frac{1}{n}[\sum_{i=1}^{n-1}(I(X\oplus x_i:z|b,s) + I(B:X\oplus x_i,z|b,s,y=i))+ \\
I(X:z|b,s) + I(B:X,z|b,s,y=0)]
\end{multline}
So that we can use Lemma 4 to the terms $\sum_{i=1}^{n-1}I(X\oplus x_i:z|b,s)+I(X:z|b,s)$ to obtain:
\begin{multline}
\frac{1}{n}[ I(X:X\oplus x_1:X\oplus x_2:..X\oplus x_{n-1}:z|b,s)+ \\ \sum_{i=1}^{n-1}I(B:X\oplus x_i,z|b,s,y=i)+ \\
 + I(B:X,z|b,s,y=0)]
\end{multline}
The terms $I(B:X\oplus x_i,z|b,s,y=i)$ are bounded by $H(B|b,s,y=i)$. Similarly the term $I(B:X,z|b,s,y=0)$ is bounded by $H(B|b,s,y=0)$ which because of the factor $\frac{1}{n}$ give rise to $H(B|b,s,y)$ and the assertion follows. $\blacksquare$ \\
Finally we come back to the proof of Theorem 2. We now proof the main result. To this end we first observe that in fact it is sufficient to show:
\begin{equation} \label{rf}
I(z:B|b,y,s) \leq \frac{1}{n}
\end{equation}
Indeed from the chain rule: $I(z:B,b,s,y)=I(z:y,b,s)+I(z:B|b,y,s)$, but $I(z:y,b,s)=0$ by assumption. Hence we get,
\begin{equation}
I(z:B,b,s,y)=I(z:B|b,y,s) \leq \frac{1}{n}
\end{equation}
which is the desired bound. To show (\ref{rf}), we use Theorem 8 and Lemma 3. From Theorem 8 we have:
\begin{multline}
 \frac{1}{n}[\sum_{i=1}^{n-1}I(X\oplus x_i:B|b,s,y=i) + I(X:B|b,s,y=0)] \\ 
I(z:B|b,s,y) \leq \frac{1}{n}I(X:X\oplus x_1:X\oplus x_2:..:X\oplus x_{n-1} |b,s) \\
+H(B|b,s,y)
\end{multline}
Now using Lemma 3 we get:
\begin{multline}
 \frac{1}{n}[\sum_{i=1}^{n-1}H(X\oplus x_i|b,s,y=i) + H(X|b,s,y=0)] \\ 
I(z:B|b,s,y) \leq \frac{1}{n}[\sum_{i=1}^{n-1}H(X\oplus x_i|b,s)+H(X|b,s)+H(z|b,s)\\-H(X,X\oplus x_1,X\oplus x_2,..X\oplus x_{n-1},z|b,s)]+H(B|b,s,y)
\end{multline}
Now because $(X|s=j)$ and $(X\oplus x_i|s=j)$ are independent from (b,y|s=j), we have for each $j$ that $H(X|b,s=j,y=0)=H(X|b,s=j)$ and $H(X\oplus x_i|b,s=j,y=i)=H(X\oplus x_i|b,s=j)$. And because for fixed $s=j$, z is independent from $b$, there is $H(z|b,s=j)=H(z|s=j)$. Averaging these equalities over $P(s=i)$ we obtain that the terms $\sum_{i=1}^{n-1}H(X\oplus x_i|b,s)+H(X|b,s)$ of LHS and RHS cancel each other and the inequality reads:
\begin{multline} 
I(z:B|b,s,y) \leq \frac{1}{n}[H(z|s)\\-H(X,X\oplus x_1,X\oplus x_2,..X\oplus x_{n-1},z|b,s)]+H(B|b,s,y)
\end{multline}
Since z is independent from s, $H(z|s)=H(z)=1$. Now, $H(X,X\oplus x_{1},X\oplus x_{2},..X\oplus x_{n-1},z|s)$ equals $H(X, x_{1},.. x_{n-1},z|s)$ as we can add $X$ to $X\oplus x_i$ reversibly. From the data processing in-equality and the independence of $s$ from $(x,z)$, we get $H(X, x_{1},.. x_{n-1},z|s)\geq H( x_{1},.. x_{n-1},z|s)=H( x_{1},.. x_{n-1},z)=n$. Hence the term  $\frac{1}{n}[H(z|s)\\-H(X,X\oplus x_1,X\oplus x_2,..X\oplus x_{n-1},z|b,s)]$ is bounded from above by $\frac{1-n}{n}$. The last term is trivially upper bounded 1, which gives desired total upper bound $\frac{1}{n}$ $\blacksquare$.
\section*{Appendix B: PROOF of lemma \ref{lem:countRAC}.}
Here we show that one requires atleast $(n-1)$ of $(2\rightarrow 1)$ RBs (or equivalently PRs) to win a $(n\rightarrow 1)$ RAC with certainty. In particular we show that $(n-2)$ $(2\rightarrow 1)$ RBs (or equivalently PRs) cannot win a $(n\rightarrow 1)$ RAC with certainty. \\

For the first step, we shall show that a no-signaling $(2\rightarrow 1)$ RB (or equivalently a PR) cannot win a $(3\rightarrow 1)$ RAC.

\begin{mydef2} \label{lem:2to13-1}
A no-signaling $(2\rightarrow 1)$ RB cannot win a $(3\rightarrow 1)$ RAC.
\end{mydef2} 
As a part of the task at hand, Alice is provided with three input bits $\tilde{a_0},\tilde{a_1},\tilde{a_2}$ and Bob with a $3$it $\tilde{b}\in \{0,1,2\}$. Additionally Alice is allowed to communicate 1 cbit of message $m$ to Bob. Finally Bob is required to guess $\tilde{B}=\tilde{a}_{\tilde{b}}$.\\
Alice and Bob share a no-signaling $(2\rightarrow 1)$  RB. Depending on $\tilde{a_0},\tilde{a_1},\tilde{a_2}$, Alice inputs $a_0\equiv a_0(\tilde{a_0},\tilde{a_1},\tilde{a_2}),a_1\equiv a_1(\tilde{a_0},\tilde{a_1},\tilde{a_2})$ to the RB. She receives an output $A$ for the RB. Alice prepares a message $m\equiv m(\tilde{a_0},\tilde{a_1},\tilde{a_2},A)$ to send to Bob. Bob inputs $b\equiv b(\tilde{b},m), A' \equiv A'(\tilde{b},m)$. He receives an output $B=a_b\oplus A \oplus A'$ from the RB . He outputs $\tilde{B}\equiv \tilde{B}(\tilde{b},m,B)$. \\
\begin{mydef4} \label{obs:1} \textit{The output of Alice's side of a no-signaling $(2\rightarrow 1)$ RB is random and uncorrelated with her inputs.} \\
Notice Bob can fix her inputs $A',b$, and $B\oplus A'=A\oplus a_b$. But under no-signaling condition Bob cannot gain any information about $a_b$. This implies that output of Alice's side of the RB $A$ must be random and generated in a way such that it is independent of her inputs $a_0,a_1$ in order to hide any information about $a_0,a_1$.
Further it follows from the fact that $a_0\equiv a_0(\tilde{a_0},\tilde{a_1},\tilde{a_2}),a_1\equiv a_1(\tilde{a_0},\tilde{a_1},\tilde{a_2})$ and $A$ is independent of $\tilde{a_0},\tilde{a_1},\tilde{a_2}$\\
\end{mydef4} 

Let us consider Bob's lab, he receives 2 bits, $m$ as message from Alice and $B$ as output from the RB given that he inputs $b=b_0$ for a particular run. 
There are the following possible strategies:
\begin{enumerate}
\item Alice sends some $m=m_0$ which depends on her inputs $\tilde{a_0},\tilde{a_1},\tilde{a_2}$ where $m_0$ is independent of $A$. Since $B=a_b\oplus A \oplus A'$ output of the RB is some random value. So without information of $A$ the RB is of no use. Now Bob is only left with one bit of information $m$. Alice does not know in advance the value $\tilde{b}$. Therefore Bob can only guess one bit with certainty. The reason for this is simply that one cannot encode more than one bit through a single bit wire reliably. To see this suppose Bob wants to learn $\tilde{a_0},\tilde{a_1}$ notice that for each value of $m$ Bob's simplest strategy can rely on the following possibilities:
\begin{enumerate}
\item $P_g(\tilde{a_0}|m=0)=1$,$P_g(\tilde{a_0}|m=1)=1$
\item $P_g(\tilde{a_1}|m=0)=1$,$P_g(\tilde{a_1}|m=1)=1$
\item $P_g(\tilde{a_0}|m=0)=1$,$P_g(\tilde{a_1}|m=1)=1$
\item $P_g(\tilde{a_1}|m=0)=1$,$P_g(\tilde{a_0}|m=1)=1$
\end{enumerate}
Notice that first two possibilities are simply sending $m=\tilde{a_0}$ and $m=\tilde{a_1}$ respectively. Further lets assume third possibility works, and Bob guess the value $\tilde{a_0}=0$ given $m=0$ and $\tilde{a_1}=0$ given $m=1$. It is easy to see that such a probability distribution $P(\tilde{a_0},\tilde{a_1},m)$ cannot exist as the probability $P(\tilde{a_0}=1,\tilde{a_1}=1)=0$. This implies the reduced distribution $P(\tilde{a_0},\tilde{a_1})$ is no longer randomly distributed, as it ought to be. Similarly arguments apply to the fourth case.

\item Alice sends $m=A$, and the RB works perfectly that is $B=a_b$. Hence depending on the choice of $b$ Bob can perfectly guess one bit in each run. W.l.o.g Alice can encode $a_0=\tilde{a_0}$ and $a_1\equiv a_1(\tilde{a_1},\tilde{a_2})$. Again as a working RB is a single bit wire given $b=1$ and it follows from observation \ref{obs:1} $A$ is uncorrelated with (has no information about) $\tilde{a_1},\tilde{a_2}$, Bob cannot perfectly guess both $\tilde{a_1}$ and $\tilde{a_2}$ simultaneously. In this case Bob can guess two bits (one for each turn, for each assignment of $b$), which is still not good enough.
\item Alice sends $m\equiv m(\tilde{a_0},\tilde{a_1},\tilde{a_2},A)$ (excluding the case $m=A$ or $m=A\oplus 1$).
Again as its a single bit, and $\tilde{a_0},\tilde{a_1},\tilde{a_2}$ are independent from $A$, Bob cannot get the value of $A$ perfectly and hence the RB wont work perfectly. Say with some probability of guessing $P_g(A|m)$ Bob could guess the value of $A$ perfectly in that case Bob can guess two bits (one for each turn) however with probability $1-P_g(A|m)$ Bob can only guess 1 bit. Therefore Bob can guess $P_g(A|m)(2) + (1-P_g(A|m))(1)=P_g(A|m)+1$ bits (one for each turn). Hence In this case Bob could guess at best two bits but the average is lower. 
 
\end{enumerate}
Therefore there exist no strategy using which Alice and Bob sharing a no-signaling $(2 \rightarrow 1)$ RB and a cbit of communication can win a $(3\rightarrow 1)$ RAC. Furthermore we make the following observation. $\blacksquare$
\begin{mydef4}\label{obs:2}
\textit{Its always better to have a working RB i.e. $m=A$ then sending a fixed message } \\
This observation directly follows from comparison between strategies one and two above. As in the case of a fixed message only one bit can be guessed with certainty while a working RB enables Bob to guess two bits (one for each turn).
\end{mydef4}
Now we proceed with proof of Lemma \ref{lem:countRAC}.
Alice is provided with $n$ input bits $\tilde{a_0},\tilde{a_1},..\tilde{a_{n-1}}$ and Bob with a $n$it $\tilde{b}\in \{0,1,..n-1\}$. Additionally Alice is allowed to communicate 1 cbit of message $m$ to Bob. Finally Bob is required to output $\tilde{B}=\tilde{a}_{\tilde{b}}$.\\

Alice and Bob share $(n-2)$ no-signaling $(2\rightarrow 1)$ RB. Depending on $\tilde{a_0},\tilde{a_1},..\tilde{a_{n-1}}$ and  outputs from other RB $A^j$ where $j\in\{1,2,..n-2\}-\{i\}$, Alice decides her inputs $a_0^i,a_1^i$ to the $i$th RB where $i\in \{1,..n-2\}$. She receives an output $A^i$ from the RB. Alice prepares a message $m$ to send to Bob depending on $\tilde{a_0},\tilde{a_1}..\tilde{a_{n-1}}$ and $A_1,A_2,..A_{n-2}$. Bob inputs $b^i, A'^i\in \{0,1\}$ to $i$th RB depending on $\tilde{b}$, output from other RB $B^j$ where $j\in\{1,2,..n-2\}-\{i\}$ and message from Alice $m$. She receives output $B^i=a_{b^i}^i\oplus A^i \oplus A'^i$ from $i$th RB.  Depending upon $B_1,B_2,..B_{n-2},m,\tilde{b}$ she outputs $\tilde{B}$. \\

Consider Bob's lab, he receives $n-1$ bits, $m$ as message from Alice and $B_i$ as output from $i$th RB given that he inputs $b^i=b_0^i$ for each run where $i\in\{1,2,..n-2\}$.

Following observation \ref{obs:2} we seek a greedy strategy, in the sense that we want to activate maximum number of RB. As there is only 1 cbit message allowed, Alice simply sends output $A_{n-2}$ in order to activate the $n-2$th RB. This allows Alice and Bob to transmit reliably two bits (though only one per run) $a_0^{n-2}$ or $a_1^{n-2}$. There are following possible strategies:
\begin{enumerate}
\item Alice uses a fixed input $a_0^{n-2}=m_0,a_1^{n-2}=m_1$ depending upon $\tilde{a_0},\tilde{a_1},..\tilde{a_{n-1}}$. In this case both the inputs $m_0,m_1$ do not carry any information about $A^1,A^2,..A^{n-3}$ so no other RB works, and $B^1,B^2,..B^{n-3}$ are random and consequently useless.  In this case as Alice is not aware of $\tilde{b}$ only 2 bits can be guessed perfectly by Bob (one for each turn depending on $b^{n-2}$). Similar arguments apply to case of Alice using inputs as functions of $\tilde{a_0},\tilde{a_1},..\tilde{a_{n-1}},A^{1},A^{2},A^{3}..A^{n-3}$ but none is exactly $A^{n-3}$.
\item Alice uses one of the available input bit to send the output of $(n-3)$th RB say $a_0^{n-2}=A^{n-3}$ and inputs $a_1^{n-2}=\tilde{a_0}$. When Bob inputs $b^{n-2}=0$, $n-3$th RB works perfectly and when $b^{n-2}=1$ Bob gets $\tilde{a_0}$ with certainty.
\item Alice uses both the available input bits to send the outputs of $(n-3)$th RB and $(n-4)$th RB and inputs $a_0^{n-2}=A^{n-3}$ and $a_1^{n-2}=A^{n-4}$. This way depending on the choice of $b^{n-2}$ an additional RB works either $n-3$th or $n-4$th. This implies $4$ bits are available for reliable communication depending on choices of Bob's input $b^{n-2},b^{n-3},b^{n-4}$. 
\end{enumerate}
Now we proceed with the same arguments as above, i.e. at each level choosing either strategy 2 or 3 and discarding 1. The strategies 2 and 3 above enforce a fully connected binary tree structure where the $(n-2)$th RB forms the root node and the rest of $(n-3)$ RB form the interior nodes. Finally free inputs form the leaf nodes. The difference between strategy 2 and 3 is that on Alice's side each RB has one input as a leaf node ($\tilde{a_i}$) and the other as output $A^i$ of $i$th RB in the former while later has both inputs as the outputs of two other RB. \\ 
We shall now present the following graph theoretic Lemma.
\begin{mydef2}
For any fully connected binary tree with a root node of degree $2$ and $k-1$  interior nodes with degree $3$, the number of leaf nodes $l$ equals $k+1$
\end{mydef2} 
For any tree the following holds,
\begin{equation}
|E|=|V|-1
\end{equation}
where $|E|$ is the number of edges, $|V|$ is the number of nodes. Let $|E|=m$, this implies,
\begin{equation} \label{subs}
m=k+l-1
\end{equation}
Also for any graph the following holds,
\begin{equation} 
2|E|=\sum_{v\in V}deg(v)
\end{equation}
so we have 1 root node with degree $2$, $k-1$ interior nodes with degree $3$ and inputs available to Alice form the leaf nodes with degree 1.  
\begin{equation}
2m=3(k-1)+l+2
\end{equation}
substituting value of $m$ from (\ref{subs}) we have,
\begin{equation}
l=k+1
\end{equation}
Hence we have number of available inputs to Alice $l$ is exactly $k+1$, where k is the number of RB available. $\blacksquare$ \\
As a part of $(n\rightarrow 1)$ RAC, Alice has $n$ input bits $\tilde{a_0},\tilde{a_1},..\tilde{a_{n-1}}$. In our case $k=n-2$, so available inputs $l=n-1$. By pigeon hole principle at least one of the available inputs must be the function of two bits provided to Alice as a part of $(n\rightarrow 1)$ RAC task. For a fixed assignment of $b^i$ this structure is a single bit wire, along with the fact $A^i$ are independent from $\tilde{a_i}$, Bob cannot reliably decode atleast one bit and hence cannot win $(n \rightarrow 1)$ RAC task with certainty. $\blacksquare$
\section*{Appendix C: $(n-1)$ no-signaling $(2\rightarrow 1)$ RBs as building blocks for $(n\rightarrow1)$ RAC.}
Here we provide a protocol for winning an $(n\rightarrow1)$ RAC using $(n-1)$ no-signaling $(2\rightarrow 1)$ RBs (or equivalently PR). We start with defining two subroutines CONCATENATION and ADDITION in general for use in the protocol for construction later. Further we use the simple fact that every natural number $n$ has a binary representation to give the protocol for the construction.   

\subsubsection*{CONCATENATION}
No-signaling $(2\rightarrow 1)$ RB can be arranged in an inverted pyramid like structure to win a $(n\rightarrow 1)$ RAC where $n=2^{k}$ for $k\in {N}$ in
the same way as the PR in the \cite{pawlowski2009information}. The trick is to
supply the outputs of the first layer of $(2\rightarrow 1)$ RB as inputs
the next layer of $(2\rightarrow1)$ RB on Alice's side. \\
\textit{For winning ($2^k \rightarrow 1$) RAC using $(2^k-1)$  no-signaling $(2\rightarrow 1)$ RB:} 
Alice has $2^{k}$ inputs bits, $a_{0},a_{1},a_{2},..a_{(2^{k}-1)}$,
she supplies them pair wise as input to $2^{k-1}$ $(2\rightarrow 1)$ RBs
which form the top most layer $r=1$ of the inverted pyramid. Each
of the no-signaling $(2\rightarrow 1)$ RBs, RB$(i)$ will give a bit output $A_{i}$ where
$i\in\{1,2,3,..2^{k-1}\}$.
Supply the outputs $A_{i}$ where $i\in\{0,1,2,..2^{k-1}\}$ pair-wise to $2^{k-2}$ no-signaling $(2\rightarrow 1)$ RBs which forms the next layer $r=2$.
Each of these RBs, RB$(i)$ will in-turn output $A_{i}$ where $i\in\{2^{k-1}+1,2^{k-1}+2,2^{k-1}+3,..2^{k-1}+2^{k-2}\}$ and repeat the above until the layer $r=k$ with $2^{k-r}=1$ $(2\rightarrow 1)$ RB($2^{k}-1$) and the final output forms the message $m=A_{2^{k}-1}$.
Bob receives $b\in\{0,1,2,3,..2^{k}-1\}$,or input bits, $b_{k}$,
which describes the index, $b=\sum_{k}b_{k}2^{k}$. Depending on which
he reads a suitable message $B_{k}$, using a box at each layer. Finally,
she outputs $B=m\oplus B_{1}\oplus B_{2}..B_{k}=a_{b}$. \\
\textit{Cost}: For $n=2^{k}$, a $(n\rightarrow1)$ RAC requires $n-1$ no-signaling $(2\rightarrow 1)$
RB. However when $n\in{N}$ and
we are only allowed to use concatenation, we first find $k\in {N}$
such that, $2^{k-1}\leq n\leq2^{k}$, then construct a $(2^{k}\rightarrow1)$
RAC using the protocol above.

For example of winning $(2^{3}\rightarrow 1)$ RAC using $(2^{3}-1)$ $(2\rightarrow 1)$ RB see Fig. 4.
\begin{figure} 
\includegraphics[scale=0.6]{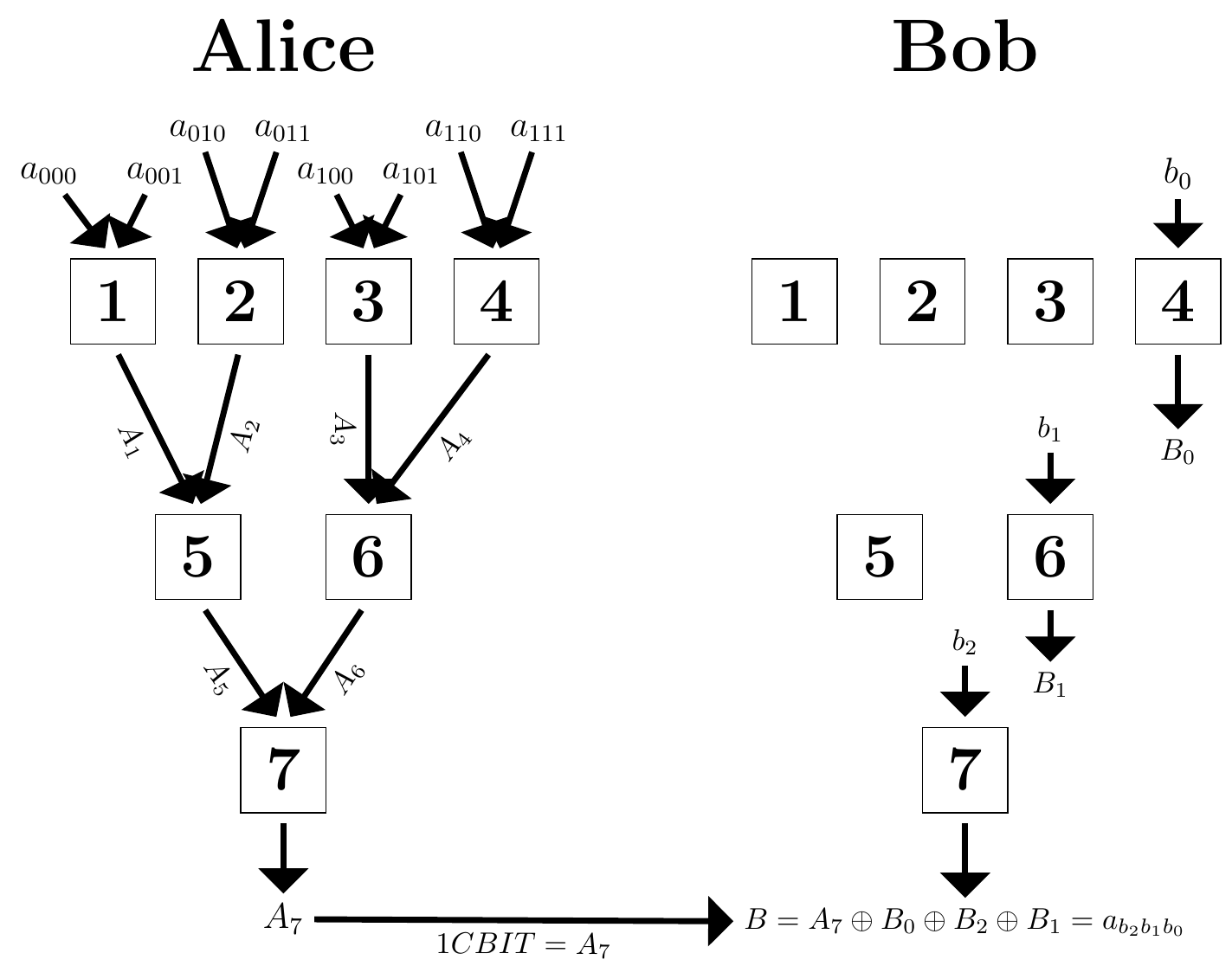}
\caption{  The figure demonstrates the use of CONCATENATION of $(2\rightarrow1)$ RB
to form a $(2^{3}\rightarrow 1)$ RB. \textit{RESOURCE: $2^{3}-1=7$ }$(2\rightarrow 1)$
RB. In the figure Bob is trying to learn $a_{101}$ or $a_{111}$.}
\end{figure}
\subsection*{ADDITION} 
Let us start with a no-signaling $(n\rightarrow 1)$ RB($1$) and no-signaling $(m\rightarrow 1)$ RB($2$).
We aim at winning $(n+m\rightarrow 1)$ RAC. The protocol to achieve that is
as follows, \\
\textit{For winning $(n+m \rightarrow 1)$ BOX using a no-signaling $(n\rightarrow 1)$ RB, no-signaling $(m\rightarrow 1)$ RB, an additional  no-signaling $(2\rightarrow 1)$ RB and 1 c-bit of communication:}
Alice has $n$ input bits $a_{0},a_{1}..a_{n-1}$ corresponding to
no-signaling $(n\rightarrow 1)$ RB($1$) and $m$ inputs bits  $a_{n},a_{n+1}..a_{n+m-1}$
corresponding to no-signaling $(m\rightarrow 1)$ RB($2$), in total $n+m$ input bits. Alice Obtains a
output bit $A_{1}$ from no-signaling $(n\rightarrow 1)$ RB($1$) and $A_{2}$ from no-signaling $(m\rightarrow 1)$
RB($2$).
The \textbf{cost} of addition is an additional no-signaling $(2\rightarrow 1)$ RB($3$) whose
input bits are $A_{1}$ and $A_{2}$ and output bit is $A_{3}$. Alice then sends $m=A_{3}$ 
On Bob's end, Bob receives $b\in\{0,1..n-1,n,..n+m-1\}$ and message
$m$. \\

If $0\leq b<n$ then Bob enters $0$ in the no-signaling $(2\rightarrow 1)$ RB(3) and
obtains $B_{3}$ and enters $b$ into no-signaling $(n\rightarrow 1)$ RB($1$) to get
$B_{1}$and outputs $B=m\oplus B_{1}\oplus B_{3}=a_{b}$.
If $n\leq b\leq n+m-1$ then Bob enters $1$ in the no-signaling $(2\rightarrow1)$ RB(3)
and obtains $B_{3}$ and enters $b$ into no-signaling $(m\rightarrow 1)$ RB($2$) to
get $B_{2}$ and outputs $B=m\oplus B_{2}\oplus B_{3}=a_{b}$.\\
For example winning $(m+n\rightarrow 1)$ RAC using no-signaling $(m\rightarrow 1)$ RB, no-signaling $(n\rightarrow1)$ RB and
a $(2\rightarrow1)$ RB [see Fig. 2].

\begin{figure} 
\includegraphics[scale=0.325]{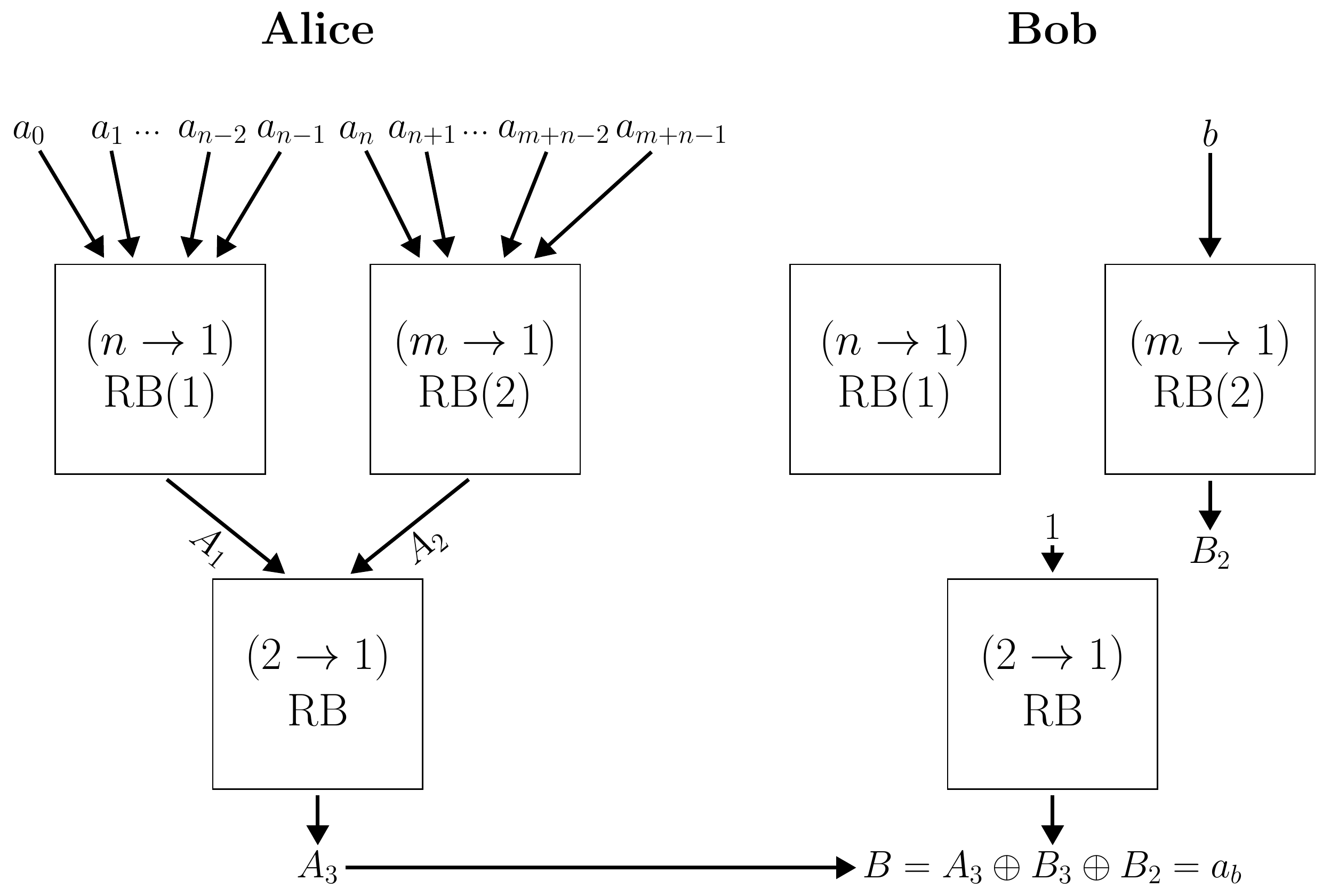}
\caption{  The figure demonstrates the use of ADDITION $(m\rightarrow1)$ RB, $(n\rightarrow 1)$ RB using
a no-signaling $(2\rightarrow1)$ RB.
RB. In the figure Bob is trying to learn $a_b$ for $n\leq b\leq n+m-1$.}
\end{figure}

Finally, protocol for winning $(n\rightarrow 1)$ RAC using only $(2\rightarrow 1)$ RB for any $n\in N$. \\
\textit{Protocol for  winning $(n\rightarrow1)$ RAC using no-signaling $(2\rightarrow1)$ RBs:} Any Natural number $n$ can be broken into sum of some discrete powers of $2$, i.e. $n=\sum_{i=0}^{k-1}\alpha_i2^i$ where $\alpha_i\in \{0,1\}$, $i\in\{0,k-2\}$ and $\alpha_{k-1}=1$, $k\in N$ such that $2^{k-1}\leq n< 2^k$. The coefficients $(\alpha_0\alpha_1..\alpha_{k-1})_2$ form the binary representation. We use a variable $count\in N$ and initialize it to $count=0$ for cost calculation given later. Alice receives $n$ input bits and for $i\in \{0,k-1\}$ such that $\alpha_i=1$ repeats the following steps,
\begin{enumerate}
\item Use CONCATENATION of $2^i-1$ no-signaling $(2\rightarrow 1)$ RB to construct a $(2^i\rightarrow 1)$ RAC for $i>0$ and for $i=0$ simply take the first input bit $a_0$. Update $count=count+1$ and RB(RIGHT)=$2^i\rightarrow 1$ RB.
\item If $count=1$ Let RB(LEFT)= RB(RIGHT).
\item  If $count>1$: use a no-signaling $(2\rightarrow1)$ RB and ADDTION of $(x\rightarrow1)$ RB(LEFT) and $(y\rightarrow1)$ RB(RIGHT) to form the updated $(x+y\rightarrow 1)$ RB(LEFT).
\end{enumerate}
Alice sends the output of final (bottom most) $(2\rightarrow1)$ RB as the message bit $m$.
Bob receives the message bit $m$ and $b\in\{0,1,2,..,n-1\}$ and outputs $B=a_b$ by following corresponding parts of CONCATENATION and ADDITION. \\
\textit{Cost}: The variable count stores the total number of indexes $i$ such that $_i=1$. CONCATENATION repeated count times uses $(n-count)$ no-signaling $(2\rightarrow1)$ RB. ADDITION is repeated $(count-1)$ each time costing $1$ no-signaling $(2\rightarrow1)$ RB. Finally total cost is $(n-count)+(count-1)=n-1$ no-signaling $(2\rightarrow1)$ RB.
For example winning $(7\rightarrow 1)$ RAC using $(6)$ $(2\rightarrow 1)$ RB [see Fig. 6].
It is to be noted at this point that the protocol given above is just one of many possibilities for simulation of $(n\rightarrow 1)$ RAC using $(n-1)$ $(2\rightarrow 1)$ RB. In particular any construction which forms a fully connected binary tree with no-signaling $(2\rightarrow 1)$ RB as nodes can be used for the simulation $(n\rightarrow 1)$ RAC. In Appendix B we give the proof that $(n-2)$ $(2\rightarrow 1)$ RB (or equivalently PR) cannot simulate $(n\rightarrow 1)$ RAC.  

\textit{Classical and quantum winning probability of $(n\rightarrow1)$ RAC using the above protocol and corresponding bounds on $(2\rightarrow1)$} RAC.\\
One can win $(n\rightarrow1)$ RAC using $n-1$ no-signaling $(2\rightarrow1)$ RB. Let the quantum winning probability over $(n\rightarrow1)$ RAC be $T_n$. As the protocol above requires no-signaling $(2\rightarrow1)$ RB, and therefore  $T_2=\frac{2+\sqrt[]{2}}{4}$ and $C_2=\frac{3}{4}$ are enough to determine $C_n$ and $T_n$ for all $n\in N$ \cite{pawlowski2009information}.  Let the winning probability of $(2\rightarrow1)$ RAC be $0<P_2<1$. As an example for $(7\rightarrow1)$ RAC [Figure 3] the probability that Bob guesses $a_b$ correctly is $P(B=a_b|b=0)=P_2^2+(1-P^2)^2$,$P(B=a_b|b=1)=(P_2^2+(1-P_2)^2)(P_2)+(1-P_2^2-(1-P_2)^2)(1-P_2)$ and $P(B=a_b|b\in \{1,2,3,4,5,6\})=P(B=a_b|b=1)$. Now Bob's inputs are uniformly distributed, hence $P_7=P(B=a_b)=\frac{P(B=a_b|b=0)+6P(B=a_b|b=1)}{7}$ So  $T=0.68723$. Similarly for $(n\rightarrow1)$ RAC for $n\in \{2,..10\}$ the protocol, $C_n$ and $T_n$ are provided in [see Fig. 7]. While $C_n$ are not optimal for $n=2k+1$ \footnote{majority function is not taken into account.}, $T_n$ are numerically close to known optimal values.

\begin{figure*} 
\includegraphics[scale=0.3]{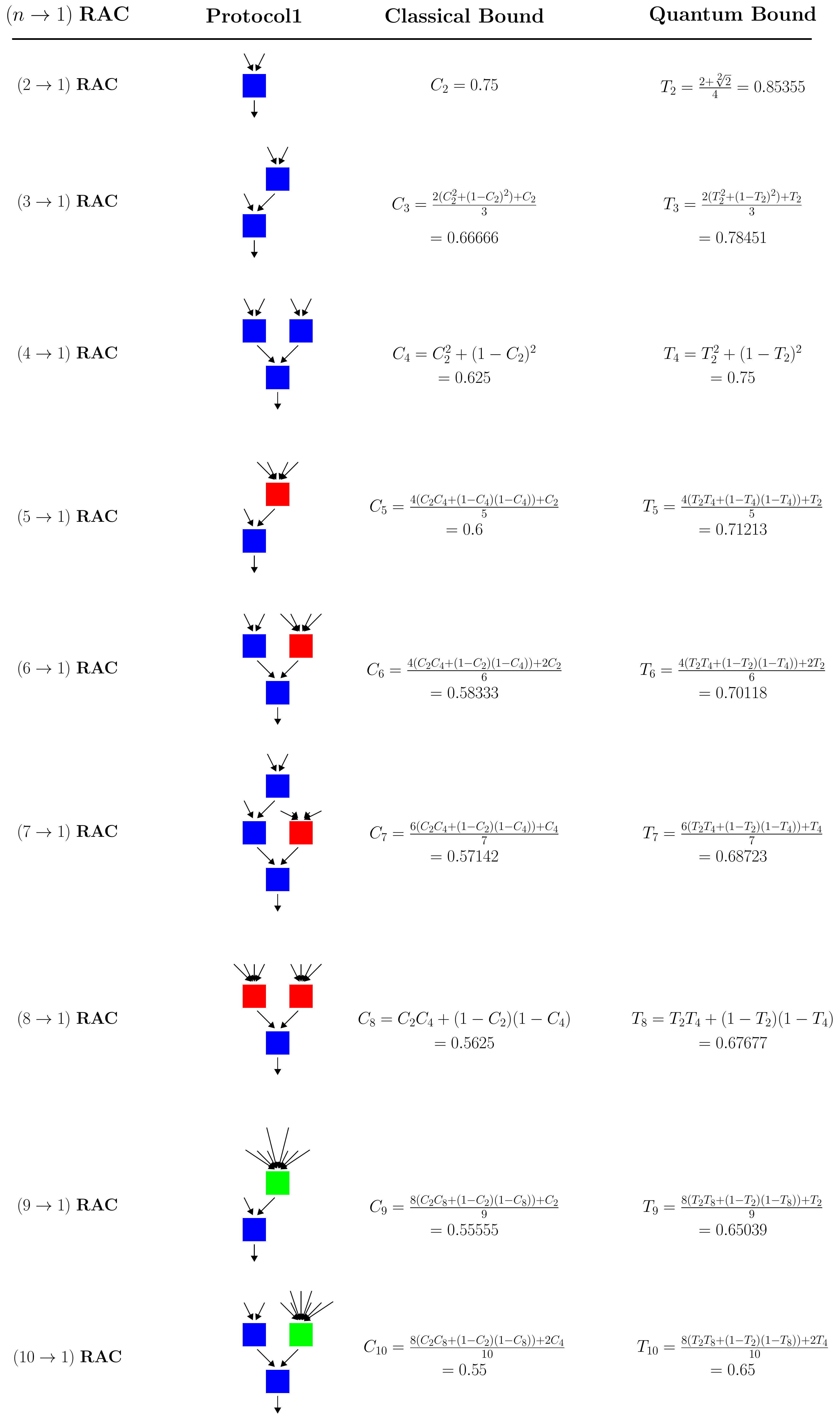}
\caption{ (Color online)  The figure demonstrates the use of protocol described in Appendix C and $n-1$ no-signaling $(2\rightarrow1)$ RB to win a $(n\rightarrow1)$ RAC for $n\in \{2,3,..10\}$. The C and Q bounds, $C_n$ and $T_n$ are derived recursively using $C_2=0.75$ and $T_2=\frac{2+\sqrt[]{2}}{4}$.  }
\end{figure*}

\section*{Appendix D}
Here we provide the proof for resource inequality \ref{re2} and consequently the in-equivalence of no-signaling $(2\rightarrow 1,3)$ RB (3) and $B^3_2(+)$. We give the proof in two parts: 
\begin{enumerate}
\item We shall show that if the $3$it of communication is not used to send the output of Alice's RB $A$ but $B^3_2(+)$-BOX is obtained, then the channel $\Lambda_3$ is a depolarizing channel: it outputs $z$  or a random $3$it with $\frac{1}{2}$ probability.
\item If the $3$it of communication is used to send $A$ and $B^3_2(+)$-BOX is obtained, then the capacity of obtainable channel $\Lambda_3$ is upper bounded by $\frac{1}{2}$ $3$it. 
\end{enumerate}
\subsection*{Proof part 1.}
Let $m$ be the message $3$it to be sent to Bob. The goal is to obtain  $B^3_2(+)$-BOX i.e. $Y=x_y-_3X$ in any case $m=0,1,2$. In general for any $m$ Bob's output is a function of his input $y$ along with the RB $Y=Y(y,b,B)$. Now there are two cases:
\begin{enumerate}
\item \textbf{$A=A'$}: In this case $B^3_2(+)$-BOX is obtained by processing a perfect $(2\rightarrow 1,3)$ RAC.
\item \textbf{$A\neq A'$}: In this case no-signaling $(2\rightarrow 1,3)$ RB (3) outputs $B$ which does not depend on the work of RAC and hence is useless. Hence Bob's outcome only depends on the input she receives i.e. $Y=Y(y)$. Now we need for  $B^3_2(+)$-BOX, $Y(y=0)=-_3X$ as $x_0=0$ and $Y(y=1)=x_1-_3X$. Notice Bob can compute $x_1$ using the fact $Y(y=1)-_3Y(y=0)=x_1$. Therefore in this case Bob must know the value of $x_1$.  
\end{enumerate}
Therefore $P_g(x_1|m=m_0)=1$ or $P_g(A|m=m_0)=1$ or both. Where $P_g$ denotes Bob's guessing probability. W.l.o.g. we assume that all three values of $m$ occur with non-zero probability. Bob's simplest strategy of (guessing only one variable, $x_1$ or $A$ for given $m$) can rely on 8 different cases:
\begin{enumerate}
\item $P_g(A|m=0)=1$,$P_g(A|m=1)=1$ and $P_g(A|m=2)=1$
\item $P_g(x_1|m=0)=1$,$P_g(x_1|m=1)=1$ and $P_g(x_1|m=2)=1$
\item $P_g(A|m=0)=1$,$P_g(A|m=1)=1$ and $P_g(x_1|m=2)=1$
\item $P_g(A|m=0)=1$,$P_g(x_1|m=1)=1$ and $P_g(A|m=2)=1$
\item $P_g(x_1|m=0)=1$,$P_g(A|m=1)=1$ and $P_g(A|m=2)=1$
\item $P_g(x_1|m=0)=1$,$P_g(x_1|m=1)=1$ and $P_g(A|m=2)=1$
\item $P_g(x_1|m=0)=1$,$P_g(A|m=1)=1$ and $P_g(x_1|m=2)=1$
\item $P_g(A|m=0)=1$,$P_g(x_1|m=1)=1$ and $P_g(x_1|m=2)=1$
\end{enumerate}
In the third case (equivalently for fourth to eighth cases) Bob makes a perfect guess of $A$ for $m=0,1$ and of $x_1$ for $m=2$. We shall now show by an example (others are analogous), that for the conditions in the third case such a joint probability distribution $P(A,x_1,m)$ cannot exist. Suppose Bob makes a perfect guess, e.g. $A=0$ for $m=0$, $A=1$ for $m=1$ and $x_1=0$. We find that $P(A=2,x_1=1)=0$ and $P(A=2,x_1=2)=0$, which implies the reduced probability distribution $P(A,x_1)$ is no longer randomly distributed, but RB works in a way such that $A$ and $x_1$ are generated independently at random.\\
From the only two possible cases we see that in the first case, $m$ is simply used to send $A$. This case is further dealt with in the Part 2. In the second case the message bit is used to send $x_1$ in this $B^3_2(+)$ BOX is obtained and the RB serves as an depolarizing $3$it channel with probability $\frac{1}{2}$. $\blacksquare$ 
\subsection*{Proof part 2}:
We shall use information theoretic tools to show that if the no-signaling $(2\rightarrow 1,3)$ RB (3) supplemented with one bit of communication is able to reproduce exactly the $B^3_2(+)$-BOX and some $3$it channel, then the mutual information of that channel must be bounded by $\frac{1}{2}$ (assuming that Alice's output of RB $A$ is directly inserted into Bob input to the RB i.e. $A=A'$). \\
We shall use the following common assumptions:\\
\textit{Assumptions:} Alice is supplied with two $3$its $x_1,z$, Bob is given a bit $y$. Both are given access to shared random $3$it $s$ such that $x_1,z,y,s$ are mutually independent. Alice generates $X$ and inputs for the RB $a_0,a_1$ from $x_1,z,s$. Similarly Bob generates his input to the RB $b$ from $y,s$. These strategies result in shared joint probability distribution $P(x_1,z,y,s,b,B,X,Y)$ such that $B=a_b$ is obtained from the RAC on Bob's side (as $A$ is always inserted into $A'$) , and $Y$ is generated out of $b,B,s,y$. 
We shall first reformulate Theorem 6 in other words.
Under the aforementioned assumptions, if variables $x_1,y,X,Y$ perfectly reproduce the $B^3_2(+)$-BOX, there holds:
\begin{equation}\label{main}
I(z:B,b,y,s)\leq \frac{1}{2}
\end{equation}

We shall prove this theorem in two parts:
\begin{enumerate}
\item First we shall use entropies and correlation to state the fact that to simulate the $B^3_2(+)$-BOX Bob has to guess perfectly $X$ when $y=0$ and $x_1-_3X$ when $y=1$. 
\item Second we shall show that it is impossible to send more than 1 $3$it through a channel with 1 $3$it capacity. As in our case Alice would like to send both $x_1$ and $z$ which bounds Bob's possible information gain about $z$. 
\end{enumerate}
\begin{mydef2}
Under the aforementioned assumptions, if variables $(x_1,y,X,Y)$ simulate the $B^3_2(+)$-BOX, there holds:
\begin{eqnarray} \label{ae1}
I(B:x_1-_3X|b,s,y=1)=\nonumber \\H(x_1-_3X|b,s,y=1) \\ \label{ae2}
I(B:X|b,s,y=0)=\nonumber \\ H(X|b,s,y=0) 
\end{eqnarray}
\end{mydef2}
Now in order to perfectly reproduce $B^3_2(+)$-BOX given $y=0$, he should perfectly guess $X$. On the other hand, given $y=1$ he should perfectly guess $x_1-_3X$. So we must have:
\begin{equation}
J(B,b,s,y=0\rightarrow X)=1 
\end{equation}
and
\begin{equation}
J(B,b,s,y=1\rightarrow x_1-_3X=1 
\end{equation}
where $J(\mathcal{X} \rightarrow \mathcal{Y})=\Sigma_iP(\mathcal{X}=i)\max_j[P(\mathcal{Y}=j|\mathcal{X}=i)]$. Which implies that there must be $max_j[P(X=j|B=l,b=l,y=0,s=i)]=1$, and consequently $H(X|B,b,y=0,s)=0$ which directly leads to (\ref{ae1}). Analogously for (\ref{ae2}). $\blacksquare$\\
\textit{One cannot send more than one $3$it through a single $3$it wire}\\
Here we prove the main argument of Theorem 6. That is, the following theorem shows the tradeoff between Bob's correlations with $X$ and $x_1-_3X$ and his correlations with $z$. 
\begin{mydef1} \label{At2}
Under aforementioned assumptions, there holds:
\begin{eqnarray} \label{pe1}
\frac{1}{2}I(x_1-_3X:B|b,s,y=1)+\nonumber \\ \frac{1}{2}I(X:B|b,s,y=1)+I(z:B|b,s,y) \leq \nonumber \\ 
\frac{1}{2}I(X:x_1-_3X:z|b,s)+H(B|b,s,y)
\end{eqnarray}
\end{mydef1}
To prove this theorem we use the following fact:
\begin{eqnarray} \label{te3}
I(S:T|V)+I(T:U|V) \leq 
I(S:U|V)+\nonumber \\ I(T:SU|V)  \equiv I(S:T:U|V)
\end{eqnarray}
The LHS of \ref{te3} can be reformulated as, upon fixing $s=i$:
\begin{eqnarray}
\frac{1}{2}I(x_1-_3X:B|b,s=i,y=1)+\nonumber \\ \frac{1}{2}I(X:B|b,s=i,y=1)+I(z:B|b,s=i,y=0) \nonumber \\ I(z:B|b,s=i,y=1)
\end{eqnarray}
Now using (\ref{te3}) to first and third terms and to second and fourth terms, we find that the above quantity is upper bounded by,
\begin{multline}
\frac{1}{2}[I(x_1-_3X:z|b,s=i,y=1)+ \\ I(B:x_1-_3,z|b,s=i,y=1)+ \\ I(X:z|b,s=i,y=0) + \\ I(B:X,z|b,s=i,y=0)]
\end{multline}
Now as $(x_1-_3X,z|s=i)$ is independent from $(y,b|s=i)$, therefore $I(x_1-_3X:z|b,s=i,y=1)=I(x_1-_3X:z|b,s=i)$. And since $(X,z|s=i)$ is independent from $(y,b|s=i)$, there is $I(X:z|b,s=i,y=0)=I(X:z|b,s=i)$. Multiplying these with $P(s=i)$ and summing over values of $s$ we obtain:
\begin{multline}
\frac{1}{2}[I(x_1-_3X:z|b,s)+ \\ I(B:x_1-_3X,z|b,s,y=1)+ \\ I(X:z|b,s) + \\ I(B:X,z|b,s,y=0)]
\end{multline}
We can now use (\ref{te3}) to the first and third terms to obtain the upper bound:
\begin{multline}
\frac{1}{2}[I(x_1-_3X:X|b,s)+ \\ I(z:X,x_1-_3X|b,s)+ \\ I(B:x_1-_3X,z|b,s,y=1) + \\ I(B:X,z|b,s,y=0)]
\end{multline}
Now $I(x_1-_3X:X|b,s)+I(z:X,x_1-_3X|b,s)=I(z:X:x_1-_3X|b,s)$. Also 
$\frac{1}{2}[I(B:x_1-_3X,z|b,s,y=1) +  I(B:X,z|b,s,y=0)]\leq H(B|b,s,y)$, and (\ref{pe1}) follows. $\blacksquare$\\
Finally, we shall proof the inequality (\ref{main}). From the chain rule: $I(z:B,b,s,y)=I(z:y,s,b)+I(z:B|y,s,b)$ but $I(z:y,s)=I(z:y,b,s)=0$ ($b=b(y,s)$). Hence \ref{main} can be reformulated as:
\begin{equation} \label{main2}
I(z:B|s,b,y)\leq \frac{1}{2}
\end{equation}
To prove (\ref{main2}) we shall reformulate (\ref{pe1}) using (\ref{ae1}) and (\ref{ae2}):
\begin{multline}
\frac{1}{2}[H(X|b,s,y=0)+H(x_1-_3X|b,s,y=1)]+\\I(z:B|b,s,y) \leq \frac{1}{2}[H(X|b,s)+H(x_1-_3X|b,s)+\\H(z|b,s)-H(X,x_1-_3X,z|b,s)]+\\H(B|b,s,y)
\end{multline}
Now as $(X|s=i)$  and $(x_1-_3X|s=i)$ are independent from $(b,y|s=i)$, we have for each $i$ that $H(X|b,s=i,y=0)=H(X|b,s=i)$ and $H(x_1-_3X|b,s=i,y=1)=H(x_1-_3X|b,s=i)$. And because for some fixed $s=i$, $z$ is independent of $b$, we have $H(z|b,s=i)=H(z|s=i)$. Averaging these equalities over $P(s=i)$ we obtain:
\begin{multline}
I(z:B|y,b,s) \leq \\ \frac{1}{2}[H(z|s)-H(X,x_1-_3X,z|b,s)]+H(B|y,s,b)
\end{multline}
Now $z$ is independent of from $s$, $H(z|s)=H(z|s)=1$. And $H(X,x_1-_3X,z|s)=H(X,x_1,z|b,s)$ as we can add $X$ to $x_1-_3X$. Using the data processing inequality and the independence of $s$ form $(x,z)$ we get, $H(z,X,x_1|s)\geq H(z,x_1|s)=H(z,x_1)=2$. Hence the first two terms are upper bounded by $-\frac{1}{2}$. The last term trivially upper bounded by $1$, which results in $\frac{1}{2}$ proving (\ref{main2}) $\blacksquare$.
\end{document}